\documentclass[useAMS,usenatbib]{mn2e}

\usepackage{graphicx}
\usepackage{color}

\newcommand{\msun}{$M_\odot$}

\newcommand{\hi}{H\,{\sc i}\rm}
\newcommand{\hii}{H\,{\sc ii}\rm}

\newcommand{\hei}{He\,{\sc i}\rm}
\newcommand{\heii}{He\,{\sc ii}\rm}
\newcommand{\siii}{[S\,{\sc iii}]}

\newcommand{\nii}{[N\,{\sc ii}]}

\newcommand{\oiii}{[O\,{\sc iii}]}
\newcommand{\oii}{[O\,{\sc ii}]}

\newcommand{\sii}{[S\,{\sc ii}]}
\newcommand{\ariii}{[Ar\,{\sc iii}]}
\newcommand{\ariv}{[Ar\,{\sc iv}]}
\newcommand{\neiii}{[Ne\,{\sc iii}]}

\newcommand{\argv}{[Ar\,{\sc v}]}

\newcommand{\op}{O$^{+}$}

\newcommand{\hp}{H$^{+}$}
\newcommand{\np}{N$^{+}$}

\newcommand{\opp}{O$^{++}$}

\newcommand{\nepp}{Ne$^{++}$}
\newcommand{\nep}{Ne$^{+}$}
\newcommand{\arpp}{Ar$^{++}$}
\newcommand{\hepp}{He$^{++}$}
\newcommand{\hep}{He$^{+}$}

\newcommand{\te}{$T_e$}

\newcommand{\linin}{$\lambda$}
\newcommand{\lin}{$\,\lambda$}
\newcommand{\llin}{$\,\lambda\lambda$}

\newcommand{\rtf}{$R_{25}$}

\newcommand{\oh}{12\,+\,log(O/H)}

\newcommand{\ohsun}{\mbox{12\,+\,log(O/H)$_\odot\,=\,$}}

\newcommand{\vs}{vs.}

\title[PNe in M33]{Planetary nebulae in M33: probes of AGB nucleosynthesis and ISM abundances\thanks{
Based on data collected at the Subaru Telescope, which is operated by the National Astronomical Observatory of Japan, and
at the W.M. Keck Observatory, which is operated as a scientific partnership among the California Institute of Technology, the University of California and the National Aeronautics and Space Administration. The Observatory was made possible by the generous financial support of the W.M. Keck Foundation.
}}
\author[F.~Bresolin et al.]{F.~Bresolin,$^{1}$\thanks{E-mail:
bresolin@ifa.hawaii.edu}
G.~Stasi\'{n}ska,$^{2}$ J.M.~V\'{\i}lchez,$^{3}$ 
J.D.~Simon,$^{4}$ and E.~Rosolowsky$^{5}$
\\
$^{1}$Institute for Astronomy, 2680 Woodlawn Drive, Honolulu, HI 96822, USA\\
$^{2}$LUTH, Observatoire de Paris, CNRS, Universit\'e Paris Diderot, Place Jules Janssen, 92190 Meudon, France\\
$^{3}$Instituto de Astrof\'{\i}sica de Andaluc\'{\i}a (CSIC), Apartado 3004, 18080 Granada, Spain\\
$^{4}$Observatories of the Carnegie Institution of Washington, Pasadena, CA 91101, USA\\
$^{5}$University of British Columbia, Okanagan, 3333 University Way, Kelowna BC V1V 1V7, Canada
}

\begin{document}



\maketitle


\begin{abstract}
\noindent
We have obtained deep optical spectrophotometry of 16 planetary nebulae in M33, mostly located in the central two kpc of the galaxy, with the Subaru and Keck telescopes. We have derived electron temperatures and chemical abundances from the detection of the \oiii\lin4363 line for the whole sample. We have found one object with an extreme nitrogen abundance, 12+log(N/H)\,=\,9.20, accompanied by a large helium content.
After combining our data with those available in the literature for PNe and \hii\ regions,
we have examined the behavior of nitrogen, neon, oxygen and argon in relation to each other, and as a function of galactocentric distance. 
We confirm the good correlation between Ne/H and O/H for PNe in M33. 
Ar/H is also found to correlate with O/H.
This strengthens the idea that at the metallicity of the bright PNe analyzed in M33, which is similar to that found in the LMC, these
elements have not been significantly modified during the dredge-up processes that take place during the AGB phase of their progenitor stars.
We find no significant oxygen abundance offset between PNe and \hii\ regions at any given galactocentric distance, despite the fact that these objects represent different age groups in the evolution of the galaxy. Combining the results from PNe and \hii\ regions, we obtain a representative slope of the ISM $\alpha$-element (O, Ar, Ne) abundance gradient in M33  of $-0.025 \pm 0.006$ dex\,kpc$^{-1}$. Both PNe and \hii\ regions display a large abundance dispersion at any given 
distance from the galactic center.
We find that the N/O ratio in PNe is enhanced, relative to the \hii\ regions, by approximately 0.8 dex.

\end{abstract}


\section{Introduction}
Planetary nebulae (PNe) can be used as test particles in the study of the chemical enrichment of galaxies and as fossil records of the nucleosynthesis
that took place in  previous generations of stars. While information on present-day chemical abundances of star-forming galaxies are gathered from the spectra of \hii\ regions and
massive stars, PNe provide the chemical makeup of older progenitor stars. Moreover,
imprinted on their emission-line spectra are the signatures of chemical elements synthesized during the AGB phase of low- and intermediate-mass stars.
Various processes occurring in these stars, such as third dredge-up and hot bottom burning, can strongly modify the helium, nitrogen and carbon abundances of the outer layers, which are eventually blown out to form the observed PNe. The effects on other elements, such as oxygen and neon, can be more subtle, and are currently the subject of lively debate (\citealt{Stasinska:2008,Cristallo:2009}).

PNe represent an alternative to \hii\ regions in the measurement of galactic abundance gradients, allowing tests
of the evolution predicted by models of galaxy assembly (e.g.~in the context of an inside-out formation, \citealt{Prantzos:2000}). In addition, PNe can be used 
to test \hii\ region abundances in cases where the latter are expected by theoretical calculations to be  affected by systematic errors. These tests are particularly important in the central, metal-rich regions of spiral galaxies, where the high metal content could lead to systematic underestimates of the nebular metallicity (\citealt{Stasinska:2005}).
This originally motivated our new investigation of a sample of luminous PNe in the central region of the galaxy M33. Despite the faintness of the targets in this external galaxy, compared to 
objects in the Milky Way, the uncertainty on the relative galactocentric distances, that afflicts studies in the Galaxy, is virtually removed. This offers a clear advantage in the
determination of abundance gradients, and consequently in deriving information on the chemical evolution of the host galaxy,
as well as providing empirical  constraints on the metallicity dependence of extragalactic distance indicators (e.g.,~the Cepheid period-luminosity relation: \citealt{Kennicutt:1998}, \citealt{Scowcroft:2009}).

In this paper we present new spectroscopic observations of 16 PNe in the neighboring galaxy M33, obtained with the aim of measuring chemical abundances from the \oiii\lin4363 diagnostic line
in the central few kpc of this galaxy.
This classical `direct' method is reckoned to yield the most accurate nebular abundances, unaffected by systematic uncertainties and/or strong-line method calibration issues, although in extragalactic PN and \hii\ region work its application is made difficult, especially at high metallicity or large redshift, by the weakness of the \lin4363 auroral line.
Combining our new data with existing observations in the literature we discuss the abundances of He, O, Ne, N and Ar in relation to each other, and as a function of galactocentric distance, in order to provide a clearer picture of the chemical composition and evolution of M33.
The number of recent studies on the present-day chemical composition of ionized nebulae (\citealt{Crockett:2006,Rosolowsky:2008}; \citealt*{Magrini:2009}) and massive stars (\citealt{U:2009}) in M33 attests to the key role played by this Milky Way neighbour in the investigation of 
the chemical evolution of spiral galaxies.

\section{Observations and data reduction}

Deep multi-slit spectra of PNe in the central region of M33 were obtained at the 8.2m Subaru Telescope equipped with the Faint Object Camera and Spectrograph (FOCAS, \citealt{Kashikawa:2002}) on 2007 October 9--10. Narrow-band \oiii\lin5007 images obtained with the same instrument were used, together with the PNe coordinates from \citet{Magrini:2001} and \citet{Ciardullo:2004},
 to select targets and prepare the multi-object masks in two circular fields of 3\arcmin\ radius. One of the fields contained the nucleus of the galaxy (the field center is approximately 1\farcm5 arcmin SE of the galaxy center), while the second one was centered 7\arcmin\ to the N of the nucleus (coordinates of the field centers are given in Table~\ref{log}).
 
For the spectroscopic observations we used 1\farcs2-wide slits and a combination of three different grisms (300R in second order, 300B and VPH650) to secure a continuous wavelength coverage from the near-UV (in most cases including the \oii\lin3727 line) to the red (up to \ariii\lin7135), with a maximum spectral resolution of 4.5\,\AA\ (300R grism in the blue) and 4.0\,\AA\ (VPH650 grism in the red). Seeing conditions during the exposures presented here varied between 0.8 and 1.0 arcsec during the first night, and between 0.5 and 0.7 arcsec during the second night. Additional spectra were obtained in 2008 under worse atmospheric conditions (seeing up to 2 arcsec), but due to their poor quality they were not included in the analysis. The spectrophotometric standards G\,191-B2B, GD\,71 and LDS\,749B were observed to flux calibrate the PNe spectra. Table~\ref{log} summarizes the exposure times used with the various grisms.

\begin{table}
 \centering
  \caption{Log of the observations (Subaru/FOCAS).}\label{log}
  \begin{tabular}{cccc}
  \hline

Field	& R.A.		& Exposure	& Grism\\
		& Dec. 	& time 		&		\\
		& (J2000) & (s)		&		\\	

 \hline
 1		& $\rm01^h33^m54\fs14$	& $3\times1800$	& 300R\\
		& $30\degr38\arcmin16\farcs5$		& $1\times1200$	& 300B\\
		&				& $2\times1200$	& VPH650 \\[1mm]		
 2		& $\rm01^h34^m03\fs61$	& $3\times1800$	& 300R\\		
		& $30\degr45\arcmin43\farcs6$		& $1\times900$	& 300B\\
		&				& $3\times1200$	& VPH650 \\	
 \hline
\end{tabular}
\end{table}

The data reduction was carried out by means of standard {\sc iraf}\footnote{{\sc iraf} is distributed by the National Optical Astronomy Observatory, which is operated by the Association of Universities for Research in Astronomy (AURA) under cooperative agreement with the National Science Foundation.} tasks, and included bias and flat-field corrections, wavelength 
calibration from thorium-argon lamp exposures, cosmic-ray removal with the {\sc l.a.cosmic} routine (\citealt{van-Dokkum:2001}), flux calibration and spectra co-addition. The subtraction of the background was carried out locally for each individual PN, from the 
flux measured within the corresponding slitlet, each of which was sufficiently long ($\sim$20 arcsec) to allow a good sampling of the background (a linear fit to the background measured in the vicinity of each PN was sufficient).
This step is crucial for the correct measurement of the low-excitation lines, which can be present in emission also in the diffuse ISM. 
It is worth pointing out in this context that in general the sky subtraction obtained from slit spectra is known to be superior to 
what can be achieved with fiber-fed spectrographs, being less prone to systematic errors.

PNe spectra were also obtained on 2004 September 22--23 and October 21
with the Keck~I telescope and the Low-Resolution Imaging Spectrometer
(LRIS, \citealt{Oke:1995}) as part of the spectroscopic survey of \hii\/ regions in the southern half of M33 by
\citet{Rosolowsky:2008}. The Keck sample extends the galactocentric distance coverage of our sample to larger radii. 
The blue spectra presented here, extending from \oii\lin3727 to \oiii\lin5007,
were obtained with the 600/4000 grism, providing 
$\sim$5\,\AA\ resolution with the 1\arcsec-wide slits used in the multi-object masks. More details on the reduction of the Keck 
dataset can be found in \citet[=\,RS08]{Rosolowsky:2008}.

\begin{figure*}
\center
\includegraphics[width=0.9\textwidth]{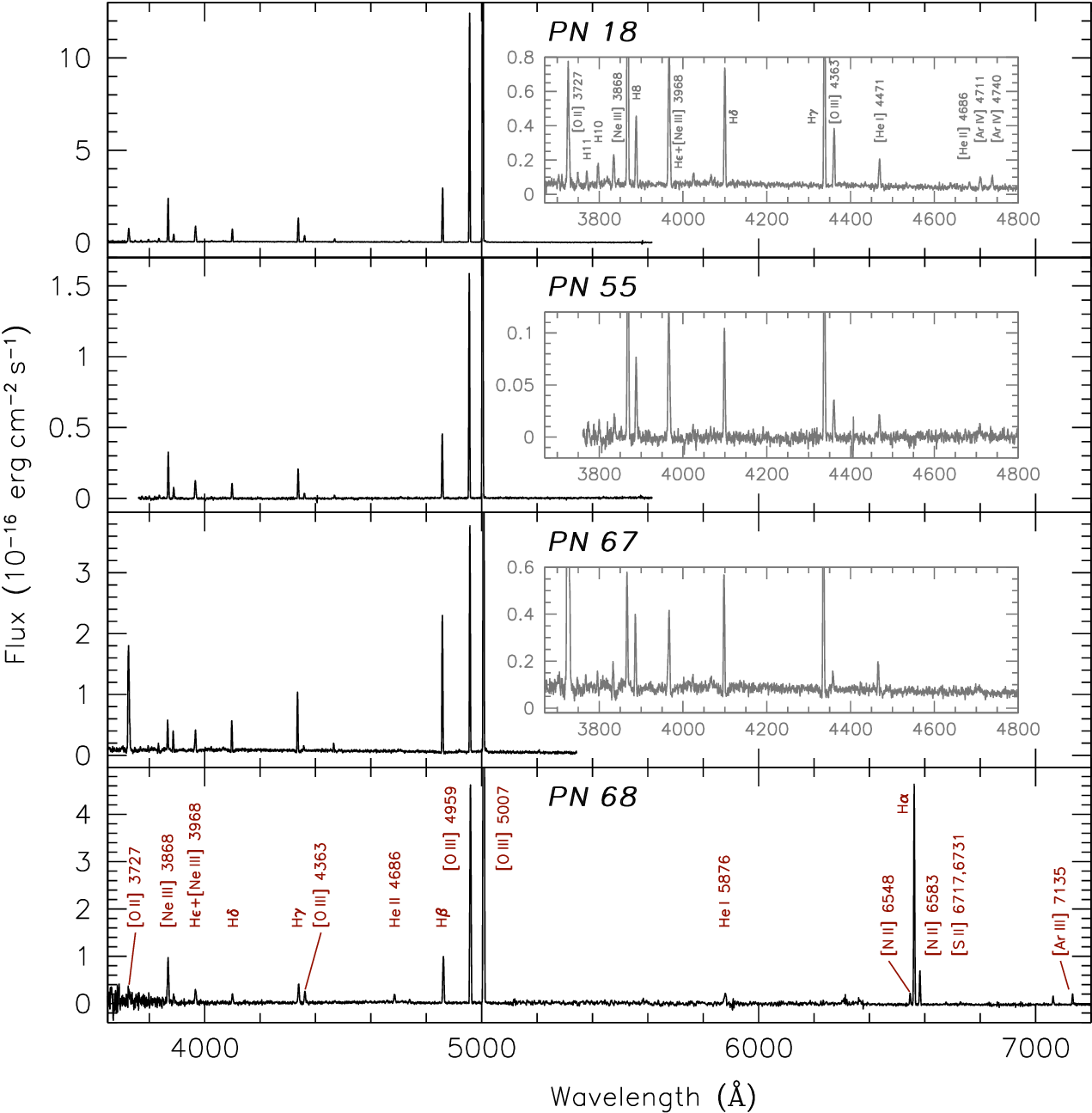}
\caption{Spectra of PN\,18, PN\,55, PN\,67 and PN\,68. Zoomed-in views are provided in the insets to better display the weakest features. The main emission lines are identified.
\label{spectra1}}
\end{figure*}

\begin{figure*}
\center
\includegraphics[width=0.9\textwidth]{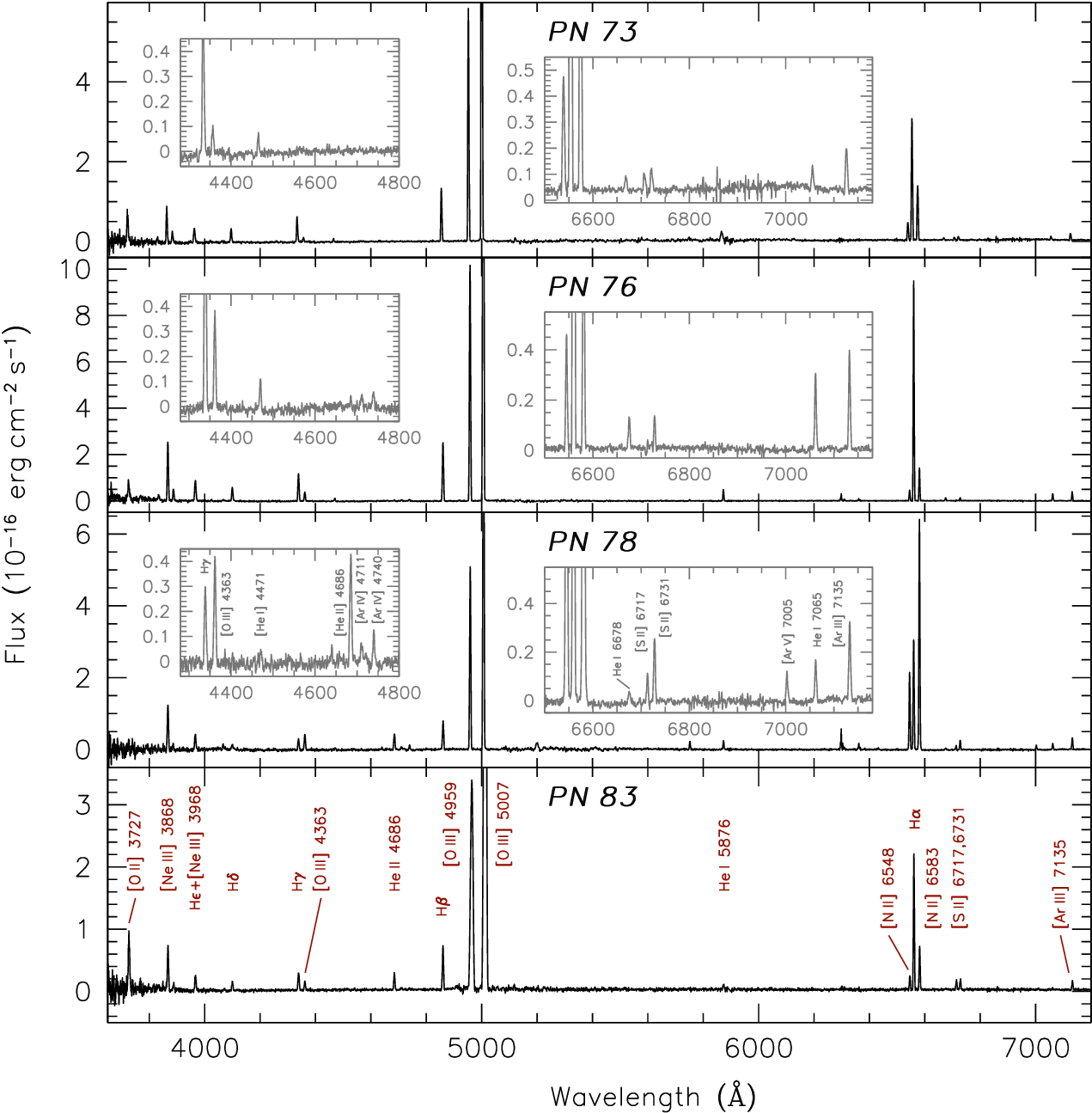}
\caption{Spectra of PN\,73, PN\,76, PN\,78 and PN\,83. Zoomed-in views are provided in the insets to better display the weakest features. The main emission lines are identified.
\label{spectra2}}
\end{figure*}

\begin{figure*}
\center
\includegraphics[width=0.9\textwidth]{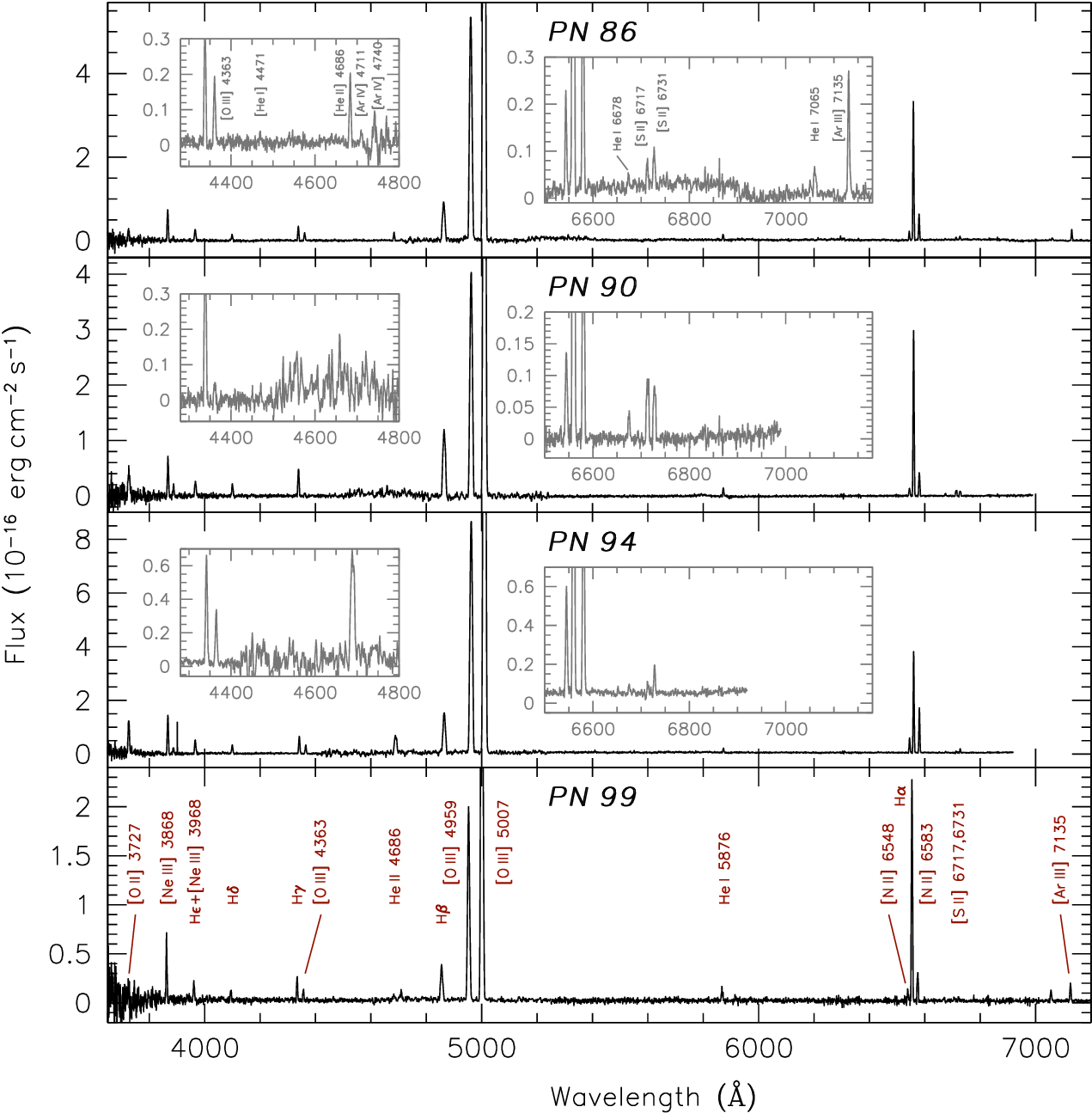}
\caption{Spectra of PN\,86, PN\,90, PN\,94 and PN\,99. Zoomed-in views are provided in the insets to better display the weakest features. The main emission lines are identified.
\label{spectra3}}
\end{figure*}

\begin{figure*}
\center
\includegraphics[width=0.9\textwidth]{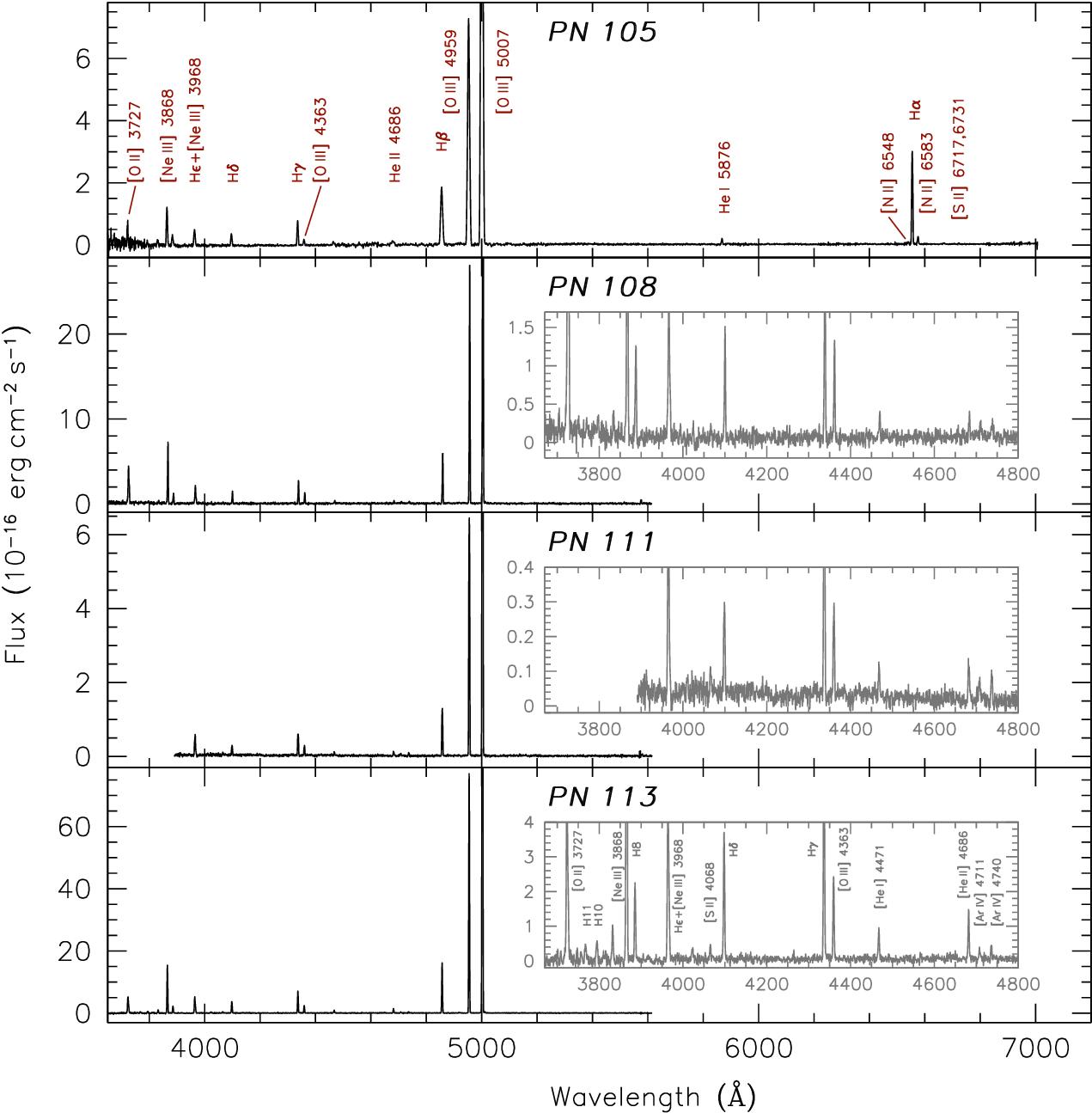}
\caption{Spectra of PN\,105, PN\,108, PN\,111 and PN\,113. Zoomed-in views are provided in the insets to better display the weakest features. The main emission lines are identified..
\label{spectra4}}
\end{figure*}

For our chemical abundance analysis we retained only spectra for which we had a reliable detection of the \te-sensitive line \oiii\lin4363.
This left us with 10 and 8 PNe from the Subaru and the Keck observations, respectively, with two targets in common between the two datasets. 
Only two Subaru targets were excluded from the analysis, because their \oiii\lin4363 line could not be measured.
The Subaru slit masks also included a few \hii\ regions. For three of them (B\,90 in Field~1, B\,72 and B\,302 in Field~2; we adopt the \hii\ region identification from \citealt{Boulesteix:1974}) we had a good quality measurement of the
electron temperature from the \oiii\lin4363 auroral lines. For the comparison between PNe and \hii\ region chemical abundances (Section~5) we retain these additional objects. For B\,72 and B\,302, to our knowledge, no previous direct abundance determinations exist in the literature.

Celestial coordinates and galactocentric distances for the final sample of PNe and \hii\ regions are summarized in Table~\ref{sample}, where the PNe coordinates and identifications are taken from the catalog by \citet{Ciardullo:2004}. For the calculation of the
de-projected galactocentric distances we adopted a disk inclination angle $i=56\degr$ and a position angle $\theta=23\degr$, 
as used by \citet{Ciardullo:2004}. The position of the galaxy nucleus was set at RA(J2000)\,=\,$\rm 01^h33^m50\fs92$, DEC(J2000)\,=\,$30\degr39\arcmin36\farcs8$ (\citealt{Massey:1996}).
For consistency with previous nebular abundance investigations, we adopted the Cepheid distance of 
840~kpc by \citet{Freedman:2001}, even though more recent studies suggest considerably larger values (up to 968~kpc: \citealt{Bonanos:2006}, \citealt{U:2009}). 
In Fig.~\ref{spectra1}-\ref{spectra4} we show the spectra of the 16 PNe included in this work. It can be seen that none of the targets has detectable 
WR-like features.

As Table~\ref{sample} shows, our sample includes 5 PNe with galactocentric distance $R \la 1$ kpc, and 11 are within 2 kpc from
the galactic center (the 3 \hii\/ regions observed by us also lie in the central 2~kpc). This central region of the galaxy is under-represented 
in emission-line studies of the chemical composition of M33, because of the observational difficulties related to the measurement 
of the faint auroral lines in metal-rich environments.
We also point out that the vast majority of the targets are expected to be part of the disk of M33.
From the measurement of the kinematics of 140 PNe \citet{Ciardullo:2004} determined 
that only two might belong to the halo of the galaxy, rather than the disk. One of them, PN\,67, is included in our sample.
Lastly, it is worth pointing out that the PNe observed as part of our program occupy the brightest 1.5 $m_{5007}$ magnitudes in the luminosity distribution of M33 planetaries.

\begin{table*}
 \centering
 \begin{minipage}{11.5cm}
  \centering
  \caption{PNe and \hii\ region sample.}\label{sample}
  \begin{tabular}{ccccccc}
  \hline

Number\phantom{aaa}	& \phantom{aa}ID$^\dag$\phantom{aa}		& \phantom{}R.A. (J2000)$^\dag$\phantom{}	& \phantom{}Dec. (J2000)$^\dag$\phantom{}					&	$R^\ddag$	&	Telescope$^\ast$	&	$c$(H$\beta$)	\\
		&		& $^h$\,\, $^m$\,\, $^s$  & $^\circ$\,\,  $'$\,\, $''$	&	(kpc)		&	&	\\

 \hline
\multicolumn{7}{c}{PNe} \\[1mm]
1\dotfill	&	PN\,18	&	01 33 06.11	&	30 31 04.5 	&	3.73		&	K	&	$0.33\pm0.08$\\
2\dotfill	&	PN\,55	&	01 33 40.16	&	30 37 49.5	&	0.88		&	K	&	$0.12\pm0.09$\\
3\dotfill	&	PN\,67	&	01 33 48.27	&	30 33 15.7 	&	1.71		&	K	&	$0.37\pm0.09$\\
4\dotfill	&	PN\,68	&	01 33 48.57	&	30 35 47.9        &	1.01		&	S	&	$0.50\pm0.13$\\
5\dotfill	&	PN\,73	&	01 33 51.06	&	30 45 38.8	&	1.70		&	S	&	$0.00\pm0.09$\\
6\dotfill	&	PN\,76	&	01 33 52.77	&	30 37 38.9	&	0.64		&	S,K	&	$0.03\pm0.08$\\
7\dotfill	&	PN\,78	&	01 33 54.68	&	30 36 05.7        &	1.17		&	S,K	&	$0.27\pm0.14$\\
8\dotfill	&	PN\,83	&	01 33 57.18	&	30 36 47.6        &	1.14		&	S	&	$0.02\pm0.11$\\
9\dotfill	&	PN\,86	&	01 33 59.97	&	30 40 28.7	&	0.74		&	S	&	$0.31\pm0.10$\\
10\dotfill	&	PN\,90	&	01 34 02.57	&	30 40 00.5	&	1.00		&	S	&	$0.16\pm0.09$\\
11\dotfill	&	PN\,94	&	01 34 03.54	&	30 39 15.6	&	1.17		&	S	&	$0.05\pm0.09$\\
12\dotfill	&	PN\,99	&	01 34 06.56	&	30 48 23.3	&	2.30		&	S	&	$0.47\pm0.13$\\
13\dotfill	&	PN\,105	&	01 34 11.91	&	30 45 00.9	&	1.87		&	S	&	$0.00\pm0.09$\\
14\dotfill	&	PN\,108	&	01 34 13.40	&	30 33 50.7 	&	3.05		&	K	&	$0.15\pm0.12$\\
15\dotfill	&	PN\,111	&	01 34 14.79	&	30 31 49.6        &	3.62		&	K	&	$0.22\pm0.10$\\ 
16\dotfill	&	PN\,113	&	01 34 15.48	&	30 32 20.3	&	3.55		&	K	&	$0.28\pm0.08$\\[2mm]
\multicolumn{7}{c}{\hii\ regions} \\[1mm]
17\dotfill	&	B\,90		&	01 34 04.34	&	30 38 12.9	&	1.40		&	S	&	$0.22\pm0.07$\\
18\dotfill	&	B\,302	&	01 34 06.92	&	30 47 26.1	&	2.09		&	S	&	$0.25\pm0.08$\\
19\dotfill	&	B\,72		&	01 34 01.21	&	30 43 56.8	&	1.20		&	S	&	$0.30\pm0.07$\\

 \hline
\end{tabular}
\end{minipage}
\begin{minipage}{11.5cm}
$^\dag$PNe coordinates from \citet{Ciardullo:2004}. \hii\ region IDs are from \citet{Boulesteix:1974}, with positions determined 
from our images.\\
$^\ddag$Galactocentric distance for $D=840$~kpc (\citealt{Freedman:2001}).\\
$^\ast$S\,=\,Subaru, K\,=\,Keck.
\end{minipage}

\end{table*}

The measurement of the emission line strengths was carried out with the {\tt splot} program in {\sc iraf}. 
The extinction $c$(H$\beta$) was derived iteratively from the Balmer decrement, assuming case B \hi\ line ratios (\citealt{Hummer:1987}) valid for the electron temperatures determined in Section~\ref{abundances}. 
The reddening-corrected line fluxes, normalized to I(H$\beta$)\,=\,100, and calculated using the \citet{Seaton:1979} interstellar reddening law, are presented in Tables~\ref{fluxes1} and \ref{fluxes2}. The extinction coefficient $c$(H$\beta$) appears in the last column of Table~\ref{sample}.
The line flux errors account for the uncertainties in the placement of the continuum, in the flat fielding and the flux calibration, and in the extinction coefficient.

In two cases (PN\,55 and 111) our spectra did not cover the bluest wavelengths, hence the entries for  \oii\lin3727 and/or  \neiii\lin3868  are missing in Table~\ref{fluxes1}. Similarly, the Keck spectra did not extend beyond \oiii\lin5007, therefore all the line fluxes at longer wavelengths are missing for the corresponding targets in Table~\ref{fluxes2}. In these tables we report the line fluxes measured independently from the Subaru/FOCAS and Keck/LRIS spectra for the two targets (PN\,76 and PN\,78) in common. This offers us a way to check the consistency 
of the flux intensities between the two samples, obtained from different instruments and reduced independently from each other.
Considering the 1-$\sigma$ errors, we find quite a good agreement between the independent measurements. 

\begin{figure}
\center
\includegraphics[width=0.45\textwidth]{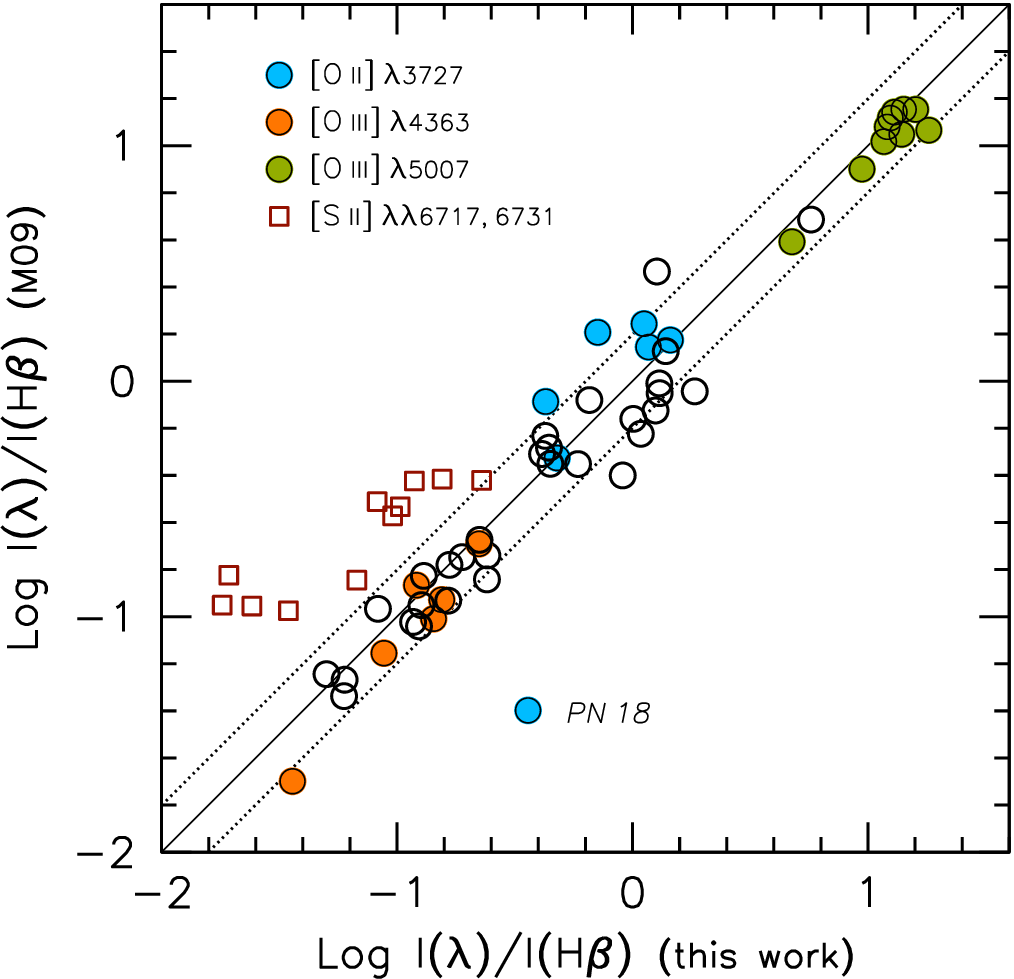}
\caption{Comparison between the emission line fluxes of various spectral lines, relative to H$\beta$, measured in this work and by \citet{Magrini:2009}, for 11 PNe in common. The line of one-to-one correspondence  is flanked by lines drawn $\pm0.2$ dex from it. Points representing O lines are shown in colour: \oii\lin3727 (blue), \oiii\lin5007 (green) and \oiii\lin4363 (orange). The outlier \oii\lin3727 line for PN\,18 is identified. The red open squares represent the 
\sii\llin6717,\,6731 lines.
 \label{comparison_flux}}
\end{figure}

We have compared our fluxes with those of \citet[=M09]{Magrini:2009} for the 11 PNe in common. The comparison is shown in 
Fig.~\ref{comparison_flux}, and it includes the O, Ne, N, S and Ar metal lines, as well as \hei\llin4471,\,5876 and \heii\lin4686.
We show in colour the points representing O lines (\oii\lin3727, \oiii\lin4363 and \oiii\lin5007), used to derive the O abundance.
The agreement is quite good for strong lines, and worsens with decreasing line strength. While most of the points
lie within the 0.2 dex scatter region from the line of equality, there are several emission features that display a larger discrepancy.
This is especially the case of the \sii\llin6717,\,6731 lines, which are often quite faint in the spectra of PNe at the distance of M33, and which appear to be systematically too strong in the M09 sample compared to ours (open red squares in Fig.~\ref{comparison_flux}). A possible explanation is that the
M09 spectra are affected by weak emission from the diffuse ionized gas (\citealt{Haffner:2009}), which we clearly detect in our 2-D spectra and which might be not properly subtracted in the work by M09, who used fibers centered in low emission regions of the galaxy for the sky subtraction. The diffuse emission would be most important for faint, low-excitation metal lines, such as those of \sii, and 
the effects of an improper subtraction on the PN spectra would increase with decreasing line strength, as observed in Fig.~\ref{comparison_flux}. 

We also found a  large systematic difference concerning the extinction coefficient, our values being lower than those reported by M09 by an average factor of 2.5. We point out that our Balmer line intensity ratios agree with the theoretical  values, while 
in several cases the H$\gamma$/H$\beta$ and H$\delta$/H$\beta$  ratios determined by M09 are significantly below the case B values.
The means we derive from their data are $0.403\pm0.102$ and $0.220\pm0.057$, respectively (case B values are 0.469 and 0.259 for \te\,=\,10$^4$~K, $N_e=10^3$\,cm$^{-3}$). This suggests a systematic over-estimate of the extinction coefficient.

The mean reddening for our PNe sample corresponds to $E(B-V)=0.19\pm0.03$. This is similar to other spectroscopic determinations of the reddening from studies of the young populations of M33, including \hii\ regions (e.g.~RS08: $E(B-V)=0.20\pm0.03$)
and young stars (\citealt{Massey:1995}: $E(B-V)=0.13\pm0.01$). The possibility that the PNe have reddening values that are  much larger 
than those measured in young star-forming regions seems unlikely, even though a somewhat higher local 
reddening could be expected for PNe, since the progenitor AGB stars can produce dust shells.

\begin{table*}
 \centering
  \begin{minipage}{16.cm}
  \centering
  \caption{Reddening-corrected fluxes (blue).}\label{fluxes1}
  \begin{tabular}{lccccccccc}
  \hline

ID	&	\oii\	&	\neiii\ 	&	H$\delta$		&	H$\gamma$	&	\oiii\	&	\hei\	&	\heii\	&	\ariv\	&	\ariv\	\\
	&	3727  	&	3868		&	4101		&	4340		&	4363		&	4471		&	4686		&	4711	 &	4740			\\

 \hline
PN\,18 &    36 $\pm$    3 &      91 $\pm$    6 &      26 $\pm$    2 &      47 $\pm$    3 &    12.2 $\pm$  0.8 &     6.0 $\pm$  3.4 &     1.1 $\pm$  0.3 &     2.8 $\pm$  0.3 &     2.6 $\pm$  0.3 \\   
PN\,55 & ...  &    73 $\pm$    5 &      25 $\pm$    2 &      47 $\pm$    3 &     8.5 $\pm$  0.8 &     5.1 $\pm$  0.8 &   (2.4) &   (2.4) &   (2.4) \\  
PN\,67 &   145 $\pm$   10 &      24 $\pm$    2 &      25 $\pm$    2 &      47 $\pm$    3 &     3.6 $\pm$  0.5 &     5.1 $\pm$  0.7 &   (0.5) &   (0.5) &   (0.5) \\  
PN\,68 &    43 $\pm$    8 &     125 $\pm$    9 &      23 $\pm$    2 &      47 $\pm$    3 &    26.9 $\pm$  2.1 &   (4.0) &    18.9 $\pm$  1.4 &   (2.6) &     5.2 $\pm$  0.3 \\   
PN\,73 &    71 $\pm$    7 &      66 $\pm$    5 &      27 $\pm$    2 &      48 $\pm$    3 &     8.8 $\pm$  0.9 &     5.4 $\pm$  0.6 &   (1.5) &   (1.5) &   (15.5) \\  
PN\,76$^\dag$ &    48 $\pm$    4 &     101 $\pm$    6 &      25 $\pm$    2 &      47 $\pm$    3 &    14.3 $\pm$  0.9 &     5.0 $\pm$  0.4 &     1.3 $\pm$  0.3 &     2.1 $\pm$  0.3 &     2.5 $\pm$  0.3 \\   
PN\,76$^\ddag$ &    41 $\pm$    3 &      86 $\pm$    6 &      24 $\pm$    2 &      47 $\pm$    3 &    13.3 $\pm$  1.1 &     4.6 $\pm$  1.8 &     2.0 $\pm$  0.5 &     2.8 $\pm$  0.5 &     3.0 $\pm$  0.5 \\   
PN\,78$^\dag$ &    24 $\pm$   14 &     184 $\pm$   13 &      26 $\pm$    3 &      47 $\pm$    3 &    58.7 $\pm$  4.0 &     8.8 $\pm$  1.7 &    58.7 $\pm$  3.7 &    13.3 $\pm$  1.7 &    14.8 $\pm$  1.7 \\   
PN\,78$^\ddag$ &    25 $\pm$    4 &     170 $\pm$   12 &      27 $\pm$    3 &      47 $\pm$    3 &    59.3 $\pm$  4.0 &     7.3 $\pm$  1.8 &    62.9 $\pm$  4.2 &    13.3 $\pm$  1.7 &    15.9 $\pm$  1.7 \\   
PN\,83 &   171 $\pm$   15 &     104 $\pm$    7 &      22 $\pm$    2 &      47 $\pm$    3 &    20.7 $\pm$  1.8 &   (3.8) &    38.7 $\pm$  2.4 &     4.9 $\pm$  0.8 &     3.8 $\pm$  0.7 \\   
PN\,86 &    47 $\pm$   10 &     113 $\pm$    8 &      26 $\pm$    2 &      47 $\pm$    3 &    26.2 $\pm$  2.0 &     3.7 $\pm$  1.2 &    24.1 $\pm$  1.6 &   (2.2) &   (3.5) \\  
PN\,90 &    69 $\pm$    6 &      68 $\pm$    5 &      23 $\pm$    2 &      47 $\pm$    3 &     8.0 $\pm$  0.8 &     5.6 $\pm$  0.9 &   (2.7) &   (2.7) &   (2.7) \\  
PN\,94 &   113 $\pm$    9 &     105 $\pm$    7 &      25 $\pm$    2 &      47 $\pm$    3 &    22.8 $\pm$  1.5 &   (3.4) &    35.4 $\pm$  2.3 &   (3.2) &   (3.2) \\  
PN\,99 &    35 $\pm$   18 &     130 $\pm$    9 &      27 $\pm$    3 &      47 $\pm$    3 &    22.9 $\pm$  2.1 &     7.0 $\pm$  1.6 &    10.6 $\pm$  1.3 &   (3.2) &     6.8 $\pm$  1.1 \\   
PN\,105 &    33 $\pm$    5 &      76 $\pm$    5 &      25 $\pm$    2 &      47 $\pm$    3 &    10.9 $\pm$  0.9 &   (1.5) &     8.3 $\pm$  0.7 &   (1.5) &   (1.5) \\  
PN\,108 &   117 $\pm$    9 &     130 $\pm$    9 &      26 $\pm$    2 &      47 $\pm$    3 &    22.3 $\pm$  1.9 &     5.8 $\pm$  1.2 &     5.3 $\pm$  1.2 &     3.1 $\pm$  1.1 &     3.4 $\pm$  1.1 \\   
PN\,111 & ...  & ...  &    23 $\pm$    2 &      47 $\pm$    3 &    20.0 $\pm$  1.7 &     6.4 $\pm$  1.2 &     9.7 $\pm$  1.2 &     6.2 $\pm$  1.2 &     6.4 $\pm$  1.2 \\   
PN\,113 &    53 $\pm$    4 &     109 $\pm$    8 &      27 $\pm$    2 &      47 $\pm$    3 &    15.5 $\pm$  1.1 &     6.0 $\pm$  0.6 &     8.3 $\pm$  0.7 &     2.2 $\pm$  0.5 &     2.6 $\pm$  0.5 \\   
B\,72 &   187 $\pm$   12 &     5.8 $\pm$  0.4 &      23 $\pm$    1 &      47 $\pm$    3 &    0.39 $\pm$ 0.04 &     4.3 $\pm$  0.3 &   (0.1) &     0.3 $\pm$ 0.04 &   (0.1) \\  
B\,90 &   196 $\pm$   12 &      71 $\pm$    4 &      24 $\pm$    1 &      47 $\pm$    3 &     5.2 $\pm$  0.4 &     6.7 $\pm$  0.4 &     4.4 $\pm$  0.3 &   (0.6) &   (0.6) \\  
B\,302 &   155 $\pm$   10 &     6.1 $\pm$  0.4 &      21 $\pm$    1 &      46 $\pm$    3 &    0.52 $\pm$ 0.04 &     3.8 $\pm$  0.2 &   (0.1) &     0.3 $\pm$ 0.03 &   (0.1) \\

 \hline
\end{tabular}

\end{minipage}
\begin{minipage}{16.cm}
Values in brackets are upper limits.\\
$^\dag$Subaru/FOCAS spectrum.
$^\ddag$Keck/LRIS spectrum.
\end{minipage}
\end{table*}

\begin{table*}
 \centering
  \begin{minipage}{15.cm}
  \centering
  \caption{Reddening-corrected fluxes (red).}\label{fluxes2}
  \begin{tabular}{lcccccccc}
  \hline

ID	&	\oiii\	&	\hei\	&	H$\alpha$	&	\nii\ 	&	\hei\	&	\sii\	&	\sii\	&	\ariii\		\\
	&	5007		&	5876  	&	6563		&	6583		&	6678		&	6717		&	6731		&	7135		\\

 \hline
PN\,18 &  1208 $\pm$   71 &   ...  & ...  & ...  & ...  & ...  & ...  & ...  \\ 
PN\,55 &  1052 $\pm$   62 &   ...  & ...  & ...  & ...  & ...  & ...  & ...  \\ 
PN\,67 &   476 $\pm$   28 &   ...  & ...  & ...  & ...  & ...  & ...  & ...  \\ 
PN\,68 &  1246 $\pm$   73 &    13.0 $\pm$  1.6 &    276 $\pm$   32 &     44 $\pm$    5 &    2.6 $\pm$  0.5 &  (2.0) &    2.9 $\pm$  0.7 &   12.7 $\pm$  1.5 \\  
PN\,73 &  1296 $\pm$   76 &    16.7 $\pm$  2.2 &    303 $\pm$   39 &    138 $\pm$   18 &    5.8 $\pm$  0.9 &    6.7 $\pm$  1.0 &    8.6 $\pm$  1.2 &   16.5 $\pm$  2.3 \\  
PN\,76$^\dag$ &  1169 $\pm$   68 &    15.2 $\pm$  1.8 &    282 $\pm$   33 &     41 $\pm$    5 &    3.8 $\pm$  0.5 &    1.8 $\pm$  0.3 &    3.5 $\pm$  0.4 &   11.7 $\pm$  1.4 \\  
PN\,76$^\ddag$ &  1186 $\pm$   70 &   ...  & ...  & ...  & ...  & ...  & ...  & ...  \\ 
PN\,78$^\dag$ &  1699 $\pm$  102 &    22.4 $\pm$  2.8 &    275 $\pm$   34 &    573 $\pm$   70 &    6.2 $\pm$  1.0 &   15.1 $\pm$  2.0 &   26.8 $\pm$  3.4 &   28.3 $\pm$  3.6 \\  
PN\,78$^\ddag$ &  1881 $\pm$  113 &   ...  & ...  & ...  & ...  & ...  & ...  & ...  \\ 
PN\,83 &  1337 $\pm$   79 &    14.4 $\pm$  2.2 &    281 $\pm$   33 &     94 $\pm$   11 &  (3.1) &   22.5 $\pm$  2.9 &   23.0 $\pm$  3.0 &   19.6 $\pm$  2.5 \\  
PN\,86 &  1734 $\pm$  103 &    13.2 $\pm$  1.9 &    288 $\pm$   34 &     53 $\pm$    6 &  (2.2) &    4.4 $\pm$  0.9 &    7.3 $\pm$  1.1 &   21.0 $\pm$  2.6 \\  
PN\,90 &   943 $\pm$   56 &    14.6 $\pm$  1.8 &    286 $\pm$   33 &     43 $\pm$    5 &    4.2 $\pm$  0.5 &   11.8 $\pm$  1.4 &   10.3 $\pm$  1.2 &  ...  \\ 
PN\,94 &  1813 $\pm$  107 &    12.4 $\pm$  1.6 &    280 $\pm$   33 &    128 $\pm$   15 &    3.4 $\pm$  0.7 &    8.2 $\pm$  1.1 &    9.6 $\pm$  1.2 &  ...  \\ 
PN\,99 &  1593 $\pm$   94 &    21.2 $\pm$  2.8 &    330 $\pm$   40 &     45 $\pm$    6 &    5.8 $\pm$  1.0 &    5.5 $\pm$  1.0 &    5.9 $\pm$  1.0 &   24.1 $\pm$  3.1 \\  
PN\,105 &  1157 $\pm$   68 &    15.8 $\pm$  2.0 &    279 $\pm$   34 &     23 $\pm$    3 &    4.1 $\pm$  0.6 &  (1.2) &    2.5 $\pm$  0.5 &  ...  \\ 
PN\,108 &  1390 $\pm$   83 &   ...  & ...  & ...  & ...  & ...  & ...  & ...  \\ 
PN\,111 &  1553 $\pm$   92 &   ...  & ...  & ...  & ...  & ...  & ...  & ...  \\ 
PN\,113 &  1418 $\pm$   83 &   ...  & ...  & ...  & ...  & ...  & ...  & ...  \\ 
B\,72 &   146 $\pm$    9 &    12.6 $\pm$  1.5 &    309 $\pm$   38 &     49 $\pm$    6 &    3.8 $\pm$  0.5 &   18.7 $\pm$  2.3 &   13.7 $\pm$  1.7 &   10.5 $\pm$  1.3 \\  
B\,90 &   698 $\pm$   41 &    14.3 $\pm$  1.7 &    280 $\pm$   33 &     42 $\pm$    5 &    3.5 $\pm$  0.4 &   34.2 $\pm$  4.0 &   24.4 $\pm$  2.9 &  ...  \\ 
B\,302 &   182 $\pm$   11 &    12.6 $\pm$  1.5 &    275 $\pm$   33 &     33 $\pm$    4 &    3.5 $\pm$  0.4 &   13.4 $\pm$  1.6 &   10.2 $\pm$  1.2 &    9.9 $\pm$  1.2 \\

 \hline
\end{tabular}

\end{minipage}
\begin{minipage}{15.cm}
Values in brackets are upper limits.\\
$^\dag$Subaru/FOCAS spectrum.
$^\ddag$Keck/LRIS spectrum.
\end{minipage}
\end{table*}

\section{Chemical abundances}\label{abundances}

 For the derivation of the ionic abundances it is necessary to obtain the electron temperature \te\ of the emitting gas.  
 In principle, different excitation zones within the nebulae are characterized
by different \te\ values, that would require the detection of  auroral lines of different ions, such as ~\oii\lin7325, \siii\lin6312 and \oiii\lin4363. 
This is observationally quite challenging for PNe at the distance of M33, and \oiii\lin4363 is the only line that we are able to use to infer \te, 
with the exception of two  \nii\lin5755 detections (PN\,76 and PN\,78).
 Therefore, for the derivation of the ionic abundances we  computed the line emissivities 
for all ions of interest  at the electron temperature obtained from the \oiii\lin4363/(\linin4959\,+\linin5007) line ratio. This includes the low-excitation line \oii\lin3727, for which a different \te\ might be appropriate,
but since
most of the oxygen in bright PNe is doubly ionized, this introduces a negligible error in the final O/H abundance.
For PNe we did not use empirical temperature schemes between high- and low-excitation zones,  since, for reasons yet not fully understood, the uncertainties would be large (see, for example, the behaviour of \te\ derived from \nii\lin5755/(\linin6548\,+\,\linin6583) as a function of \te\ from \oiii\lin4363/(\linin4959\,+\linin5007) for a sample of Galactic PNe in \citealt{Gorny:2009}). For \hii\ regions, however, we derived the temperature of the low-excitation zone (for the \op\ and \np\ ionic abundances) from the \citet{Garnett:1992} relation, T(\op)\,=\,0.7\,T(\opp) + 3000 K,
which is supported by recent observations of extragalactic \hii\ regions (\citealt{Esteban:2009, Bresolin:2009a}).

\begin{table}
 \centering
  \caption{Sources of atomic data.}\label{atomic}
  \begin{tabular}{ccc}
  \hline

Ion		&	Transition		&		Collision \\
		&	probabilities	&		strengths\\
\hline
\oii			& Froese Fischer \&								& \citet{Tayal:2007} 			\\
			& Tachiev (2004)\nocite{Froese-Fischer:2004}				&												\\[1mm]
			
\oiii		& Froese Fischer \&								& \citet{Aggarwal:1999} 			\\
			& Tachiev (2004)				&												\\[1mm]
			
\nii			& Froese Fischer \&								& \citet{Hudson:2005} 			\\
			& Tachiev (2004)				&												\\[1mm]

\neiii		& Froese Fischer \&								& \citet{McLaughlin:2000} 			\\
			& Tachiev (2004)				&												\\[1mm]

\ariii		& \citet{Mendoza:1983}		& \citet{Galavis:1995} 								\\
			& Kaufman \&\nocite{Kaufman:1986}		& 													\\
			& Sugar~(1986)				&			\\[1mm]

\sii			& Froese Fischer 		& \citet{Ramsbottom:1996} 							\\
			& et al.~(2006)\nocite{Froese-Fischer:2006}			&				\\[1mm]

\siii		& Froese Fischer			& \citet{Tayal:1999} 			\\	
			& et al.~(2006)			&				\\

 \hline
\end{tabular}
\end{table}

The electron temperatures were derived  using the five-level program {\em temden} (\citealt*{De-Robertis:1987}), in the implementation of {\sc iraf}'s {\em nebular} package (\citealt{Shaw:1995}).  The adopted atomic parameters are the same as in \citet{Bresolin:2009a}, and are summarized in Table~\ref{atomic} for convenience. For \hei\ we used the emissivities of \citet*{Porter:2007}, including the correction for collisional excitation, and for \hi\ and \heii\ those 
of \citet{Hummer:1987}.

The electron densities were estimated from the \sii\lin6717/\linin6731 ratios, when available (Subaru data only), or
from \ariv\lin4711/\linin4740, if measured. We arbitrarily assumed $N_e=1000$ cm\,$^{-3}$ in the remaining cases. The density affects the \te\ determination only weakly, even though it has an important impact on the \op\ emissivity.
For example, around 12,000~K, a variation of $N_e$ between 1000\,cm$^{-3}$ and 5000\,cm$^{-3}$
induces a change in T\oiii\ of 60\,K, which is on the order of 10\% of the estimated temperature errors. The emissivity of the 
\oii\lin3727 line changes by about 50\%, but we have verified that this has little effect on our results.
For the final \te\ uncertainties 
(rounded to the nearest hundred K in Table~\ref{te-ne}) we have propagated the 1-$\sigma$ errors in the line fluxes (contained in Tables~\ref{fluxes1}-\ref{fluxes2}), and include the 
small contribution from the uncertainties in $N_e$. 
Table~\ref{te-ne} summarizes 
the plasma diagnostics thus obtained, and the density values that we adopted. 
We note the generally good agreement between densities obtained from the \sii\ and \ariv\ line ratios, when both are available. In these cases (PN\,76, PN\,78 and PN\,83) we have adopted the \sii-based densities for the following analysis, because of their smaller uncertainties. Table~\ref{te-ne} reports the plasma diagnostics 
for the two targets in common between the Subaru and Keck datasets
(PN\,76 and PN\,78).
Within the errors, we find a good agreement between independent determinations for  both the T\oiii\ electron temperatures
and the \ariv-based electron densities.

\begin{table*}
 \centering
  \begin{minipage}{12.5cm}
  \centering
  \caption{Plasma diagnostics.}\label{te-ne}
  \begin{tabular}{lccccc}
  \hline

\phantom{aaaaa}ID\phantom{aaaaa}	&	T\oiii	&	T\nii	&	$N_e$\,(\sii)		&	$N_e$\,(\ariv)	& $N_e$ (adopted)\\
\hline
PN\,18\dotfill & 11500 $\pm$  500 &   ...              &      ...              &       3290 $\pm$   820  &    3290 $\pm$   820 \\ 
PN\,55\dotfill & 10700 $\pm$  500 &   ...              &      ...              &      ...              &       1000 $\pm$   100 \\ 
PN\,67\dotfill & 10400 $\pm$  600 &   ...              &      ...              &      ...              &       1000 $\pm$   100 \\ 
PN\,68\dotfill & 15700 $\pm$ 1000 &   ...              &      ...              &      ...              &       1000 $\pm$   100 \\ 
PN\,73\dotfill & 10100 $\pm$  500 &   ...              &       1060 $\pm$   120  &   ...              &       1060 $\pm$   120 \\ 
PN\,76$^\dag$\dotfill & 12300 $\pm$  600 &   13500 $\pm$ 3200 &       6020 $\pm$  1420  &    8320 $\pm$  1520  &    6020 $\pm$  1420 \\ 
PN\,76$^\ddag$\dotfill & 11900 $\pm$  600 &   ...              &      ...              &       6050 $\pm$  1540  &    6050 $\pm$  1540 \\ 
PN\,78$^\dag$\dotfill & 20100 $\pm$ 1600 &   14700 $\pm$ 2500 &       4910 $\pm$   380  &    7680 $\pm$  1210  &    4910 $\pm$   380 \\ 
PN\,78$^\ddag$\dotfill & 18900 $\pm$ 1400 &   ...              &      ...              &       9360 $\pm$  1310  &    9360 $\pm$  1310 \\ 
PN\,83\dotfill & 13500 $\pm$  800 &   ...              &        540 $\pm$    50  &    1100 $\pm$   990  &     540 $\pm$    50 \\ 
PN\,86\dotfill & 13400 $\pm$  800 &   ...              &       2980 $\pm$   840  &   ...              &       2980 $\pm$   840 \\ 
PN\,90\dotfill & 10900 $\pm$  600 &   ...              &        240 $\pm$    20  &   ...              &        240 $\pm$    20 \\ 
PN\,94\dotfill & 12500 $\pm$  600 &   ...              &        850 $\pm$    80  &   ...              &        850 $\pm$    80 \\ 
PN\,99\dotfill & 13200 $\pm$  800 &   ...              &        660 $\pm$   150  &   ...              &        660 $\pm$   150 \\ 
PN\,105\dotfill & 11200 $\pm$  500 &   ...              &      ...              &      ...              &       1000 $\pm$   100 \\ 
PN\,108\dotfill & 13600 $\pm$  900 &   ...              &      ...              &       6310 $\pm$  3360  &    6310 $\pm$  3360 \\ 
PN\,111\dotfill & 12500 $\pm$  700 &   ...              &      ...              &       5230 $\pm$  1530  &    5230 $\pm$  1530 \\ 
PN\,113\dotfill & 11700 $\pm$  600 &   ...              &      ...              &       7180 $\pm$  2150  &    7180 $\pm$  2150 \\ 
B\,72\dotfill &  7800 $\pm$  300 &   ...              &         51 $\pm$     5  &   ...              &         51 $\pm$     5 \\ 
B\,90\dotfill & 10400 $\pm$  400 &   10200 $\pm$ 1400 &         22 $\pm$     2  &   ...              &         22 $\pm$     2 \\ 
B\,302\dotfill &  8000 $\pm$  300 &    8100 $\pm$  700 &         79 $\pm$     8  &   ...              &         79 $\pm$     8 \\

 \hline
\end{tabular}

\end{minipage}
\begin{minipage}{12.5cm}
$^\dag$Subaru/FOCAS spectrum.
$^\ddag$Keck/LRIS spectrum.
\end{minipage}

\end{table*}

The abundances of \op, \opp, \nepp, \np, \arpp\, \hep\ and \hepp, relative to \hp, were derived with the {\em ionic} program in {\sc iraf}, from the 
de-reddened line fluxes and the \te, $N_e$ values found in the previous step. 
Total chemical abundances were obtained adopting the ionization correction factors (ICFs) of \citet{Kingsburgh:1994}, with the exception of argon for the \hii\ regions, in which case we adopted the recipes by \citet{Izotov:2006}.
The ionization correction factor for oxygen is obtained as (1\,+\,He$^{2+}$/He$^+$)$^{2/3}$, where the He$^+$ abundance has been derived 
from \hei\lin5876 when available (Subaru dataset), and from \hei\lin4471 in the remaining cases.
Our final abundances are shown in Table~\ref{abundancetable}, where the uncertainties have been 
calculated by propagation of the errors in the line fluxes and \te. Once again, we point out the good agreement in the resulting abundances between independent measurements for PN\,76 and PN\,78, observed both with Subaru and Keck.    The weighted abundances for these two PNe (also included in 
Table~\ref{abundancetable}) will be adopted for the remainder of the paper.

For two of the Keck targets, PN\,55 and PN\,111, the O$^+$ abundance could not be calculated, because the \oii\lin3727 line was outside the covered wavelength range, therefore the quoted value is a lower limit to the O/H ratio. However, the median O$^+$/O ratio of the remaining sample is 0.06, exceeding 0.12 only in the case of PN\,67 (0.26), so the total oxygen abundance in PN\,55 and PN\,111 is likely to be close to the values reported
in Table~\ref{abundancetable}.

\begin{figure}
\center
\includegraphics[width=0.47\textwidth]{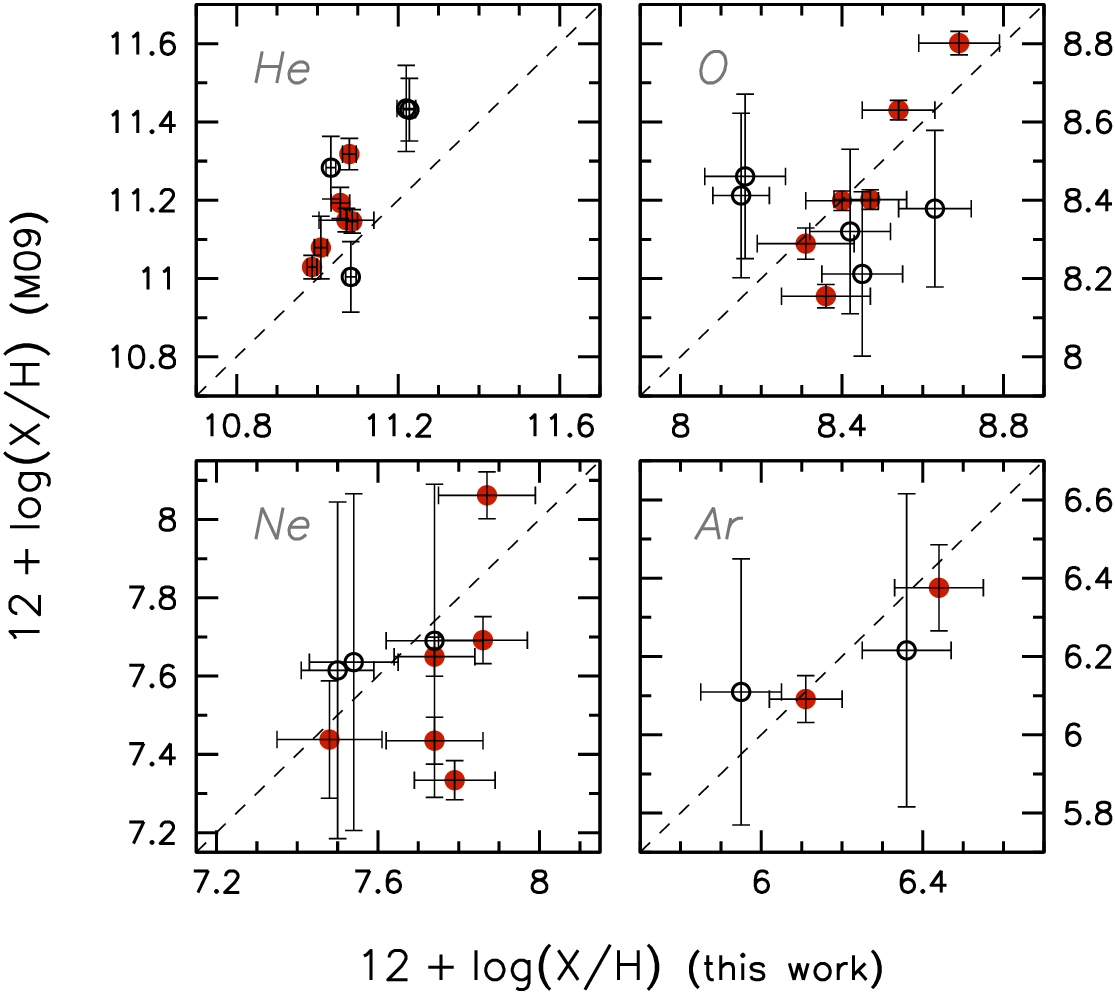}
\caption{Comparison between the chemical abundances of He, O, Ne and Ar  measured in this work and by \citet[=M09]{Magrini:2009}  for PNe in common.  Dashed lines are drawn to represent the one-to-one correspondence. Objects with an electron temperature determination based on \oiii\lin4363 from the M09 sample are shown as red full dots.
\label{comparison_abund}}
\end{figure}

In Fig.~\ref{comparison_abund} we compare our final PNe abundances with those obtained by M09. 
These authors have determined electron temperatures either from direct detections of auroral lines (\oiii\lin4363, \nii\lin5755), or from a statistical relationship between \te\ and the strength of \heii\lin4686. Abundances obtained
with the latter method are not expected to be accurate, neither on theoretical grounds, nor on the basis of the diagrams by M09 which were used to define this relation. In order to differentiate the two \te-determination methods, we represent 
the \oiii\lin4363 detections in the M09 sample by full points in Fig.~\ref{comparison_abund}, while open symbols are used for PNe without a \lin4363 line detection (we remind the reader that
\oiii\lin4363 is measured for each PN contained in our sample).
It can be seen that the agreement between our abundances and those by M09 is generally
acceptable, and that no obvious systematic offset exists between the two sets of
abundances,  with the exception of helium. This is likely a result of different abundance calculations, in particular 
regarding the effect of collisional excitation, because the \hei\ line fluxes are in good agreement between the two studies. The remaining large discrepancies ($>$\,0.2 dex) in Fig.~\ref{comparison_abund} are due to significant differences either in the measured diagnostic line strengths
(as in the case of Ne), or in the adopted \te\ (in the case of the most discrepant points for O, which do not have a \oiii\lin4633-based \te\ determination in M09). The plot for O confirms the expectation that PNe in the M09 sample without an \oiii\lin4363\ detection, and whose abundances derive from a  \te\ determination based on \heii\lin4686,
are affected by larger random errors than those with an observed \oiii\lin4363 line, and that a large abundance scatter
is expected for these objects.

One of the three \hii\ regions we observed, B\,90, was also included in the abundance studies by \citet{Crockett:2006} and RS08.
Both obtained \oh\,=\,8.50$\pm$0.06, in very good agreement with our value \oh\,=\,8.47$\pm$0.08. The Ne/H abundance ratios are also virtually coincident.

\medskip
For the remainder of the paper, we 
complement our dataset of 16 PNe, covering mostly the inner region of M33, with the PNe sample studied by M09, but considering only objects with a direct determination of the oxygen abundance, i.e.~based on the measurement of the \oiii\lin4363 auroral line. This restricts the number of PNe from M09 to 32 (out of 93), extending in galactocentric distances from 0.6 to 8 kpc. Of these, 6 are in common with our sample: PN\,18, PN\,67, PN\,73, PN\,76, PN\,108, and PN\,113.
New auroral line-abundances for 10 PNe are therefore added by our study, 7 of them at $R<1.2$~kpc. For consistency, we have recalculated the chemical abundances from the line fluxes published by M09, with the same procedure used for our PNe sample.
In the plots that follow, in the cases where the M09 and our sample overlap, we have retained only the abundance information
obtained from our observations, in order to avoid representing the same object more than once.

 As mentioned above,  \citet{Ciardullo:2004} identified two PNe (PN\,24 and PN\,67) as possibly belonging to the M33 halo, 
having velocities that deviate more than 2.5\,$\sigma$ from the mean rotation curve. However, they 
point out that these outliers can be expected out of the total number of 140 targets, simply on statistical grounds.
We find that the chemical abundances
of PN\,67 are comparable with those of the other targets in our sample. PNe in the halo of the Milky Way are generally
metal-poor compared to those in the disk (e.g.~\citealt*{Howard:1997,Henry:2004}). We conclude that PN\,67 likely belongs to the disk of M33, rather than the halo.
However, we also point out that PN\,67 is the only object in our PN sample that clearly shows absorption components in the 
Balmer lines, an indication of the presence of an underlying stellar component, as in the case of \hii\ regions. It has also the lowest excitation ratio observed in our sample, O$^{++}$/O\,=\,0.74, which overlaps with the range observed for \hii\ regions in M33 (however, PNe with even lower excitation are known, e.g.~in the Milky Way, see \citealt{Kingsburgh:1994} or \citealt{Gorny:2009}). PN\,67
is not among the brightest in H$\alpha$ flux or in $m_{5007}$ (\citealt{Ciardullo:2004}).
It is therefore unclear whether this is truly a PN, or a compact, high-excitation \hii\ region.
For the remainder of this paper we will continue to include PN\,67 along the remaining PNe. None of the results presented in the next sections suggests that we should do otherwise.

\begin{table*}
 \centering
  \begin{minipage}{13cm}
  \centering
  \caption{Chemical abundances.}\label{abundancetable}
  \begin{tabular}{lccccc}
  \hline

\phantom{aaaaa}ID\phantom{aaaaa}	&	He/H		&	12\,+\,log(O/H)		&	12\,+\,log(N/H)		&	12\,+\,log(Ne/H)		&	12\,+\,log(Ar/H)		\\
\hline
PN\,18\dotfill &  0.12 $\pm$  0.07 &   8.47 $\pm$  0.09  &  ...                & 7.79 $\pm$  0.10  &  ...                \\ 
PN\,55\dotfill &  0.10 $\pm$  0.02 &  8.49$^*$$\pm$ 0.10  &  ...                & 7.79 $\pm$  0.11  &  ...                \\ 
PN\,67\dotfill &  0.10 $\pm$  0.02 &   8.31 $\pm$  0.12  &  ...                & 7.48 $\pm$  0.13  &  ...                \\ 
PN\,68\dotfill &  0.11 $\pm$  0.01 &   8.16 $\pm$  0.10  &   8.07 $\pm$  0.11  &   7.54 $\pm$  0.11  &   5.95 $\pm$  0.10    \\ 
PN\,73\dotfill &  0.12 $\pm$  0.02 &   8.69 $\pm$  0.10  &   8.62 $\pm$  0.12  &   7.87 $\pm$  0.12  &   6.44 $\pm$  0.11    \\ 
PN\,76$^\dag$\dotfill &  0.10 $\pm$  0.01 &   8.38 $\pm$  0.09  &   7.89 $\pm$  0.10  &   7.75 $\pm$  0.10  &   6.11 $\pm$  0.09    \\ 
PN\,76$^\ddag$\dotfill &  0.09 $\pm$  0.04 &   8.43 $\pm$  0.10  &  ...                & 7.72 $\pm$  0.11  &  ...                \\ 
PN\,76$^+$\dotfill &  0.10 $\pm$  0.01 &   8.40 $\pm$  0.09  &   7.89 $\pm$  0.10  &   7.74 $\pm$  0.10  &   6.11 $\pm$  0.09    \\ 
PN\,78$^\dag$\dotfill &  0.17 $\pm$  0.02 &   8.11 $\pm$  0.10  &   9.32 $\pm$  0.11  &   7.48 $\pm$  0.11  &   6.13 $\pm$  0.11    \\ 
PN\,78$^\ddag$\dotfill &  0.18 $\pm$  0.03 &   8.21 $\pm$  0.10  &  ...                & 7.51 $\pm$  0.11  &  ...                \\ 
PN\,78$^+$\dotfill &  0.17 $\pm$  0.01 &   8.15 $\pm$  0.07  &   9.32 $\pm$  0.11  &   7.50 $\pm$  0.09  &   6.13 $\pm$  0.11    \\ 
PN\,83\dotfill &  0.14 $\pm$  0.02 &   8.41 $\pm$  0.10  &   8.01 $\pm$  0.11  &   7.71 $\pm$  0.11  &   6.25 $\pm$  0.11    \\ 
PN\,86\dotfill &  0.11 $\pm$  0.01 &   8.49 $\pm$  0.10  &   8.27 $\pm$  0.11  &   7.71 $\pm$  0.11  &   6.29 $\pm$  0.10    \\ 
PN\,90\dotfill &  0.11 $\pm$  0.02 &   8.45 $\pm$  0.10  &   7.99 $\pm$  0.11  &   7.77 $\pm$  0.12  &  ...                \\ 
PN\,94\dotfill &  0.12 $\pm$  0.01 &   8.63 $\pm$  0.09  &   8.47 $\pm$  0.10  &   7.81 $\pm$  0.10  &  ...                \\ 
PN\,99\dotfill &  0.17 $\pm$  0.02 &   8.42 $\pm$  0.10  &   8.36 $\pm$  0.11  &   7.74 $\pm$  0.12  &   6.36 $\pm$  0.11    \\ 
PN\,105\dotfill &  0.12 $\pm$  0.02 &   8.49 $\pm$  0.09  &   8.04 $\pm$  0.11  &   7.76 $\pm$  0.11  &  ...                \\ 
PN\,108\dotfill &  0.11 $\pm$  0.02 &   8.36 $\pm$  0.11  &  ...                & 7.74 $\pm$  0.12  &  ...                \\ 
PN\,111\dotfill &  0.13 $\pm$  0.02 &  8.47$^*$$\pm$ 0.10  &  ...                & 7.74 $\pm$  0.12  &  ...                \\ 
PN\,113\dotfill &  0.12 $\pm$  0.01 &   8.54 $\pm$  0.09  &  ...                & 7.86 $\pm$  0.11  &  ...                \\ 
B\,72\dotfill &  0.09 $\pm$  0.01 &   8.46 $\pm$  0.09  &   7.45 $\pm$  0.09  &   7.62 $\pm$  0.12  &   6.31 $\pm$  0.11    \\ 
B\,90\dotfill &  0.11 $\pm$  0.01 &   8.47 $\pm$  0.08  &   7.53 $\pm$  0.08  &   7.94 $\pm$  0.10  &  ...                \\ 
B\,302\dotfill &  0.09 $\pm$  0.01 &   8.44 $\pm$  0.08  &   7.36 $\pm$  0.08  &   7.52 $\pm$  0.10  &   6.25 $\pm$  0.10    \\

 \hline
\end{tabular}

\end{minipage}
\begin{minipage}{13cm}
$^*$Lower limit.\\
$^\dag$Subaru/FOCAS spectrum.
$^\ddag$Keck/LRIS spectrum.\\
$^+$Adopted abundances (weighted mean of Subaru and Keck results).
\end{minipage}

\end{table*}

\section{Excitation properties}\label{results}

\begin{figure*}
\includegraphics[width=0.95\textwidth]{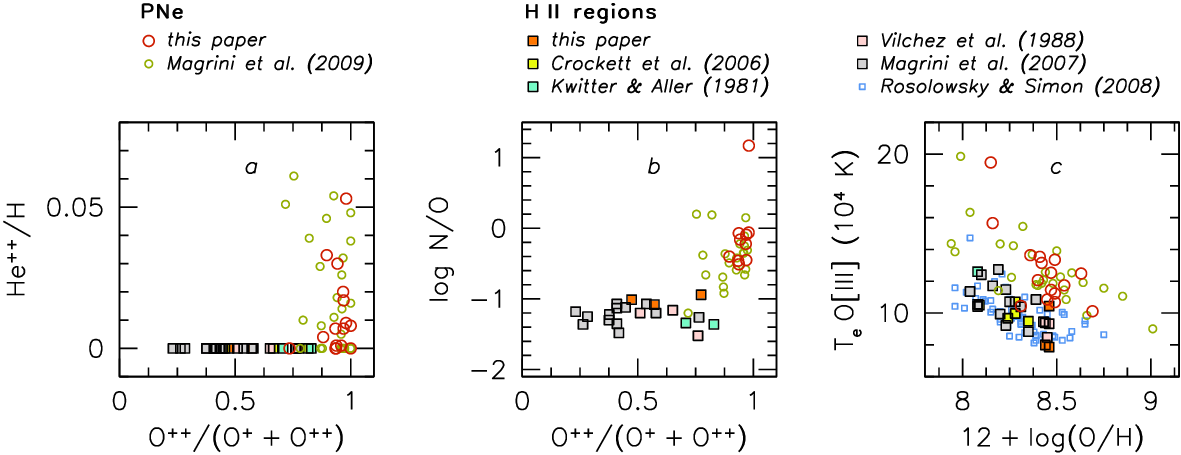}
 \caption{ (a) \hepp/H abundance ratio  as a function of the nebular excitation
 \opp/(\op + \opp). (b) N/O abundance ratio  as a function of the nebular excitation.
 (c) electron temperature  as a function of O/H.
 The objects plotted are: our new sample of PNe (red open circles), PNe in \citet{Magrini:2009} with a \oiii\lin4363 detection (green open circles), \hii\ regions with auroral line-based abundances from a compilation of data in the literature  and
 the current work (squares). The sources for the data are summarized in the legend at the top.
  \label{excitation}
}
\end{figure*}

A first look at the main properties of the PNe in M33 is presented in Fig.~\ref{excitation}, 
where we show the \hepp/H and N/O abundance ratios  as a function of the excitation \opp/(\op\,+\,\opp) (panels a and b).
In this and subsequent plots we use red open circles to represent our new PN sample, while the green circles represent the PNe observed by M09 and with a \oiii\lin4363 line detection. We also show with orange squares our new \hii\ region abundances (B\,72, B\,90 and B\,302),
and with  the remaining squares  a sample of \hii\ regions for which auroral line-based abundances could be derived, using line fluxes we compiled from the literature (\citealt{Kwitter:1981,Vilchez:1988, Crockett:2006, Magrini:2007}; RS08). For consistency, we recomputed the PN and \hii\ region abundances adopting the same techniques explained in Sect.~3.

Fig.~\ref{excitation} highlights
the fact that the excitation of PNe is very high, with more than 75\% of the oxygen in the \opp\ ionization stage for most of the objects shown.
The \hii\ regions can be clearly separated from the majority of the PNe in Fig.~\ref{excitation}a simply from their \hepp/H ratios,  since \heii\lin4686, when observed in \hii\ regions, never exceeds the intensity of a few percent of H$\beta$, while in planetary nebulae this line is often observed and can reach intensities similar to H$\beta$. 
The separation between PNe and \hii\ regions is well displayed also by looking at N/O \vs\ \opp/(\op\,+\,\opp) in Fig.~\ref{excitation}b. 
The average N/O ratio of the PNe is obviously larger than that for the \hii\ regions, a result of the 
nucleosynthesis processes taking place in the advanced stages of evolution of the PNe stellar progenitors. This is also reflected by the large dispersion seen for the PN data, compared to the \hii\ regions.
Finally, in Fig.~\ref{excitation}c we plot the ionic temperature \te\oiii, derived from the \oiii\lin4363/(\linin4959\,+\,\linin5007) line ratio, 
as a function of the oxygen abundance. The general trend, for both PNe and \hii\ regions, is of decreasing electron temperature with increasing metallicity. 
However, at any given O/H ratio, the PNe have higher \te\ values, as expected, and display a larger dispersion, which 
reflects the dispersion in stellar effective temperatures.

\subsubsection*{Uncertainties in the ICFs?}

\begin{figure}
\center
\includegraphics[width=0.47\textwidth]{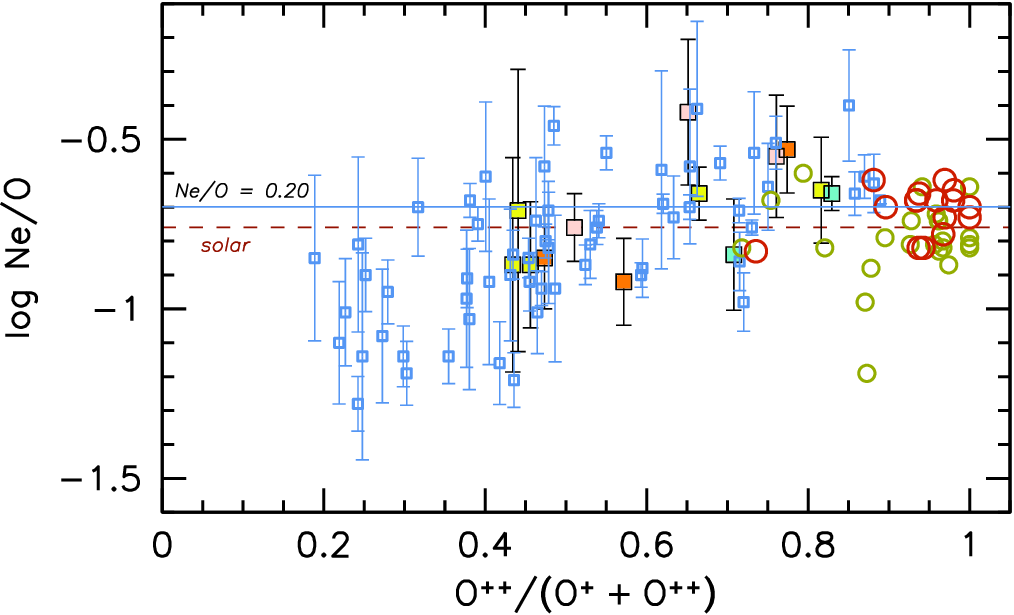}
\caption{The Ne/O ratio as a function of the excitation ratio \opp/(\op+\opp).  Symbols as in Fig.~\ref{excitation}.
The error bars for the \hii\ regions are shown. The horizontal dashed line represents the solar value log(Ne/O)$_\odot$\,=\,$-0.76$ (\citealt{Asplund:2009}).  The mean Ne/O\,=\,0.20 for the  PNe in our sample is shown by the continuous blue line.
\label{ne_excitation}
}
\end{figure}

Fig.~\ref{ne_excitation} shows how the \hii\ region Ne/O ratio appears to decrease with excitation, in contrast with the expectation that this ratio should remain constant with variations of the nebular ionization structure. Since our neon abundances are derived from the single \neiii\lin3868 line, this result can originate from an inaccurate 
 correction for unseen ionization stages, in particular \nep, when the \opp/(\op + \opp) ratio is below 0.5. 
 However, it is unclear whether the discrepancy at 
\opp/(\opp+\op)\,$<$\,0.5 is due to problems with the ICF(Ne) or to systematic errors in the data, because some of the low-excitation objects 
have a `normal', approximately solar-like Ne/O ratio [log\,(Ne/O)$_\odot$\,=\,$-0.76\pm0.11$, \citealt{Asplund:2009}, shown by the horizontal dashed line]. Moreover, some of the high-excitation  \hii\ regions significantly exceed both the solar value 
and the PN values.
We have considered alternative schemes for the ICF(Ne) of \hii\ regions, such as those proposed by \citet{Izotov:2006} and \citet{Perez-Montero:2007}, but found  even larger  trends with excitation. We have therefore continued to adopt the ICF(Ne) from \citet{Kingsburgh:1994} for the rest of this paper, in order to minimize the excitation dependence of the Ne/O ratio.

Considering only our PN sample, we obtain a mean $<$Ne/O$>$\,=\,$0.20\pm0.03$, shown by the continuous line  in 
Fig.~\ref{ne_excitation}. We suggest to use this value as the representative Ne/O ratio in the ISM of M33, because the selected PNe are the objects 
with the highest excitation (and presumably with the most accurate abundances), and therefore the uncertainty in the total Ne abundance due to corrections for \nep\ should be negligible.

\begin{figure}
\center
\includegraphics[width=0.47\textwidth]{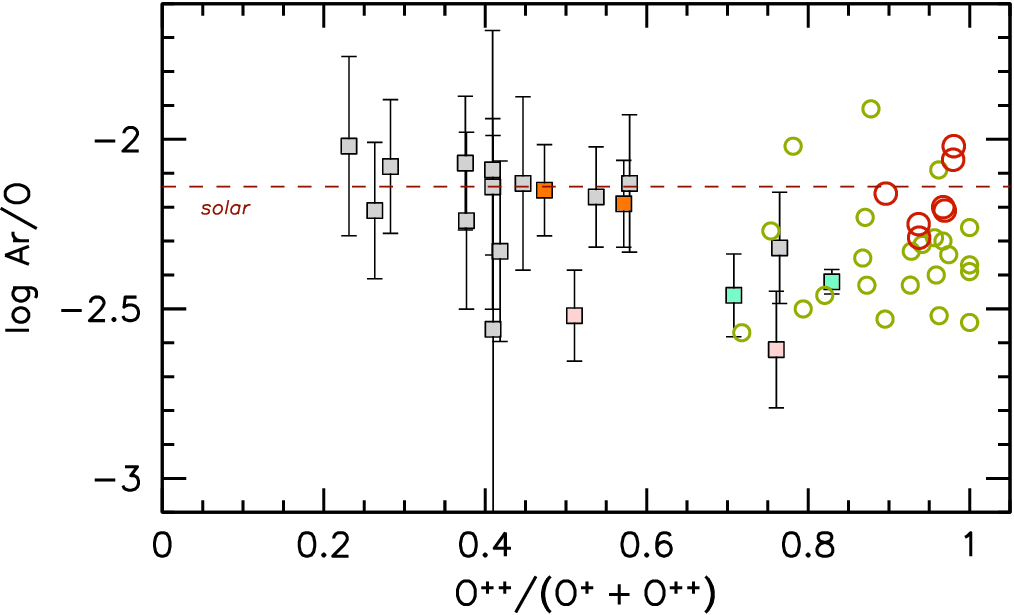}
\caption{The Ar/O ratio as a function of the excitation ratio \opp/(\op\,+\,\opp).  Symbols as in Fig.~\ref{excitation}.
The error bars for the \hii\ region data points are shown. The horizontal dashed line represents the solar value log(Ar/O)$_\odot$\,=\,$-2.14$ (\citealt{Lodders:2003}).
\label{ar_excitation}
}
\end{figure}

As we explained earlier, for argon we have adopted the ICFs from \citet{Kingsburgh:1994} for PNe and from \citet{Izotov:2006} for \hii\ regions.
 A plot of the Ar/O abundance as a function of the excitation \opp/(\opp\,+\,\op) is shown in Fig.~\ref{ar_excitation}. As in the case of neon, the \hii\ regions show a trend, this time with high-excitation objects having a lower Ar/O ratio than lower-excitation ones. The latter
agree with the solar ratio, log(Ar/O)$_\odot$\,=\,$-2.14\pm0.09$  (\citealt{Lodders:2003}, indicated by the horizontal dashed line in Fig.~\ref{ar_excitation}). This suggests that the ICF(Ar) we are adopting is inadequate at \opp/(\opp\,+\,\op)\,$>$\,0.6.
 We also point out the existence of a systematic offset of the Ar/O ratio between our PN sample (red circles), which agrees with the solar value, and the one based on the M09 data, which has a lower mean Ar/O. This could be related to observational errors, or could be due to the systematically higher extinction found  by M09 (which would lower the \ariii\lin7135 strength relative to H$\beta$). For the following discussion we must therefore keep in mind that high-excitation \hii\ regions and the M09 sample of PNe appear to have argon abundances that are under-estimated.

\section{Abundance patterns}

\subsection{Nitrogen and helium}

\begin{figure}
\includegraphics[width=0.47\textwidth]{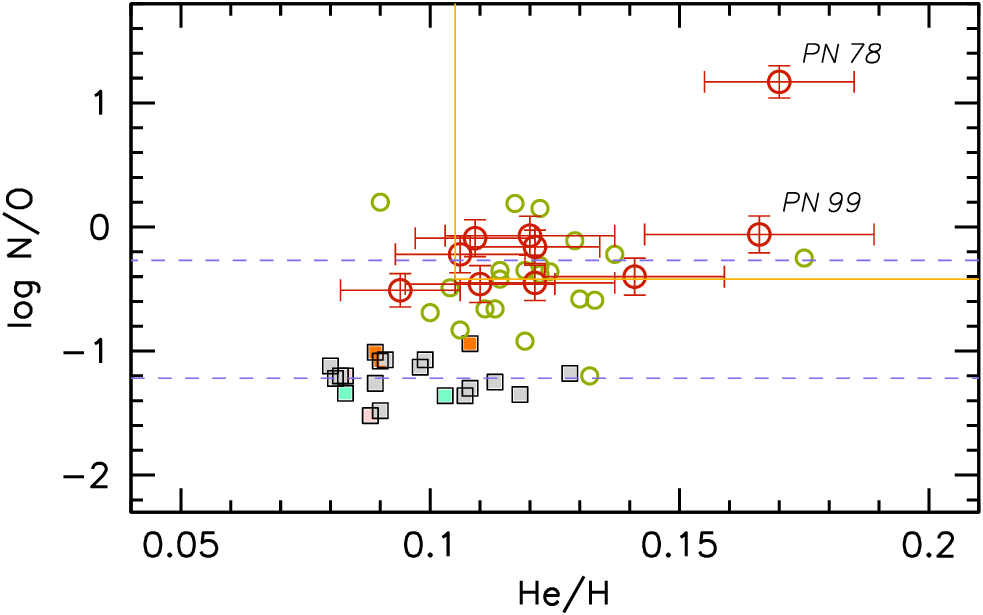}
 \caption{The N/O abundance ratio as a function of the He/H ratio. The boundaries defined by \citet{Torres-Peimbert:1997} for Type~I PNe in the LMC (N/O~$>$~0.38, He/H~$>$~0.105)  are drawn in orange. Symbols as in Fig.~\ref{excitation}. Error bars are shown for our sample.
 The mean N/O ratio is drawn with dashed horizontal lines for the \hii\ regions (lower line) and our PN sample (upper line).
  \label{nhe}
}
\end{figure}

The impact of the nucleosynthetic processes occurring in the PN progenitors on the chemical compositions of the ejected envelopes can be visualized in a plot of N/O \vs\ He/H (Fig.~\ref{nhe}). Both N and He can be enhanced at the stellar surface during the red giant and AGB phases, largely depending on the mass and metallicity of the stellar precursors. The details are complicated by the combined effects of mechanisms such as mass loss, rotation, hot bottom burning and third dredge-up (\citealt{Charbonnel:2005}).
The comparison between our new PN sample (mean log(N/O)\,=\,$-0.27$, excluding PN\,78) and the \hii\ region data (square symbols, with a mean log(N/O)\,=\,$-1.22$) shows a N/O enhancement on the order of one dex  (the two mean values are shown as dashed lines in Fig.~\ref{nhe}). When we factor in the radial trends (Sec.~\ref{gradient_nitrogen}), the typical enhancement in N/O is $\sim0.8$ dex.

Stars belonging to the high end of the PN progenitor mass spectrum  (which extends approximately between 0.8 and 8~\msun)  are associated with Type~I PNe (\citealt{Peimbert:1978}), and are part of the youngest stellar population, showing clear N and He abundance excesses relative to the composition of \hii\ regions, which represent the chemical makeup of the present-day ISM. The values of the minimum N/O and He/H ratios that define this class of PNe vary slightly between different authors, and need to account for the value of the N and O abundances in the ISM out of which the progenitor stars were formed (e.g.~\citealt*{Stasinska:1998,Leisy:1996}).
Since the metal content of M33 is comparable to that of the LMC,  we show in Fig.~\ref{nhe} the corresponding limits, N/O~$>$~0.38, He/H~$>$~0.105, proposed by \citet{Torres-Peimbert:1997}. With this definition, most of the PNe in our sample (red symbols with error bars)
are of Type~I, or have an N/O ratio that is compatible, given the errors, with being of Type~I. However, it is worth pointing out that most of the PNe in our sample
have N/O ratios that are only slightly above the \citet{Torres-Peimbert:1997} limit for inclusion into the Type~I definition.
In fact, we do not expect that a large number of PN progenitors in our sample, which is drawn from the brightest PNe in M33, experienced
evolutionary phases that can strongly enhance the N/O ratio, such as hot bottom burning or second dredge-up, because the corresponding  progenitor masses would be around 2.5\,\msun, according to the models of \citet{Marigo:2004}. Furthermore, the IMF would favor the detection of lower mass progenitors with respect to high mass ones.

\subsubsection{The remarkable PN\,78}

The largest helium content in our PN sample (He/H\,$\simeq$\,0.17) is reached for PN\,78 and PN\,99 (both are identified in Fig.~\ref{nhe}). The former also has a striking
nitrogen excess, 12\,+\,log(N/H)\,=\,9.20 and log(N/O)\,=\,1.09. 
The large N content does not depend on the 
correction factor scheme that we used, and is due to the strong nitrogen lines observed (the strength of \nii\lin6583 is approximately twice that of H$\alpha$; see the spectrum in Fig.~\ref{spectra2}). The high electron temperature is accompanied by a very high nebular excitation, with the only detection of \argv\llin6435,\,7005 in our sample.

\begin{figure}
\includegraphics[width=0.47\textwidth]{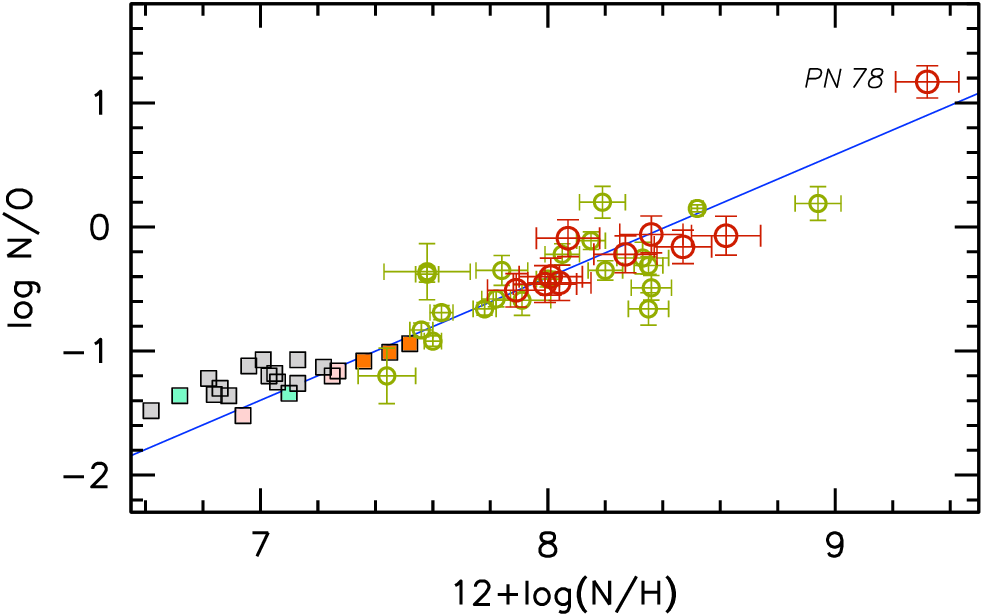}
 \caption{The N/O abundance ratio as a function of the N/H ratio. Symbols as in Fig.~\ref{excitation}. 
 The regression to the PN data is represented by the continuous line.
   \label{nohh}
}
\end{figure}

The nitrogen enrichment  that we derive for PN\,78 is comparable to the maximum observed to date in Type~I PNe belonging to the Milky Way
(e.g.~\citealt{Stanghellini:2006}), although it occurs at a much lower O/H ratio. In fact, PN\,78 has the lowest measured 
oxygen abundance in our sample, \oh\,=\,8.15, considerably lower than the median of 8.47. It is tempting to interpret 
these data as the result of hot bottom burning in a massive progenitor ($\sim 5\,M_\odot$), which could explain both the low O and the high N abundances, with nitrogen synthesized at the expense of C. At sufficiently high temperatures N could be produced at the expense of oxygen in the ON cycle.  However, in the case of PN\,78 the latter possibility seems to be excluded. We plot in Fig.~\ref{nohh}  the N/O ratio as a function of N/H. The regression line to the whole PN sample (excluding PN\,78), calculated accounting for errors in both coordinates (we used
an IDL adaptation of the routine {\em fitexy} by \citealt{Press:1992} for this and subsequent fits of this kind), has a slope of $0.99\pm0.01$, indicating that the N/O enhancement relative to \hii\ regions is due primarily to high N production, rather than O depletion. This appears to hold
for the M33 PN sample as a whole, and also for the specific case of PN\,78.
In addition, both the Ar/O and Ne/O ratios for PN\,78 are comparable to those of the rest of the sample, confirming that oxygen is not significantly depleted.

PNe observed in the Milky Way and the Magellanic Clouds have generally much smaller N/H ratios than we find in PN\,78, which
is interpreted as a modest efficiency of hot bottom burning in intermediate-mass precursors (\citealt{Marigo:2003}), since a larger efficiency would lead to an overproduction of N. Our data for PN\,78 seem to suggest that, 
occasionally, large N/H and N/O ratios can be encountered, simultaneously with a large He/H ratio (the latter as a result of an efficient third dredge-up).

\citet{Richer:2007} have presented chemical abundances for PNe in the irregular galaxy NGC~6822. One of their targets, S33, with 12\,+\,log(N/H)\,=\,8.93 and log(N/O)\,=\,0.90, has a nitrogen excess similar to PN\,78 in M33. In this case, however, the oxygen abundance, \oh\,=\,8.03, is {\em larger} than the galaxy's mean value. These authors propose that this could be an example of a star in which oxygen has been dredged-up in the late evolutionary phases. The high N/O ratio is not explicitly discussed by \citet{Richer:2007}, although they note that, in general, nitrogen should be more
easily dredged-up than oxygen.

\subsection{Neon and oxygen}\label{neon}
Since core-collapse supernovae (massive stars with initial $M>8\,M_\odot$) are the main sites of oxygen and neon production, one expects that, unless important nucleosynthesis of these elements takes place in the less massive stars that are the PN precursors, their variation occurs in lockstep in PNe, similarly to the case of \hii\ regions. 
This is in fact observed in the Milky Way and the Magellanic Clouds  (\citealt{Henry:1989, Leisy:2006}),
and is normally taken as evidence that the oxygen abundance measured in PNe is equivalent to that of the progenitor stars at the moment of birth (\citealt{Richer:2007}). Deviations from this result have been discovered in low metallicity environments, where third dredge-up episodes could be more efficient, with the consequence that oxygen can be overabundant in PNe relative to \hii\ regions, as both theory (\citealt{Marigo:2001}) and observations (\citealt{Magrini:2005a}; \citealt*{Pena:2007,Kniazev:2008}) suggest. However, according to \citet{Richer:2007} these findings constitute the exception rather than the rule,
perhaps because most progenitors of the bright PNe studied spectroscopically in other galaxies are of rather small mass ($<1.5\,M_\odot$, \citealt{Richer:2008}).

\begin{figure}
\includegraphics[width=0.47\textwidth]{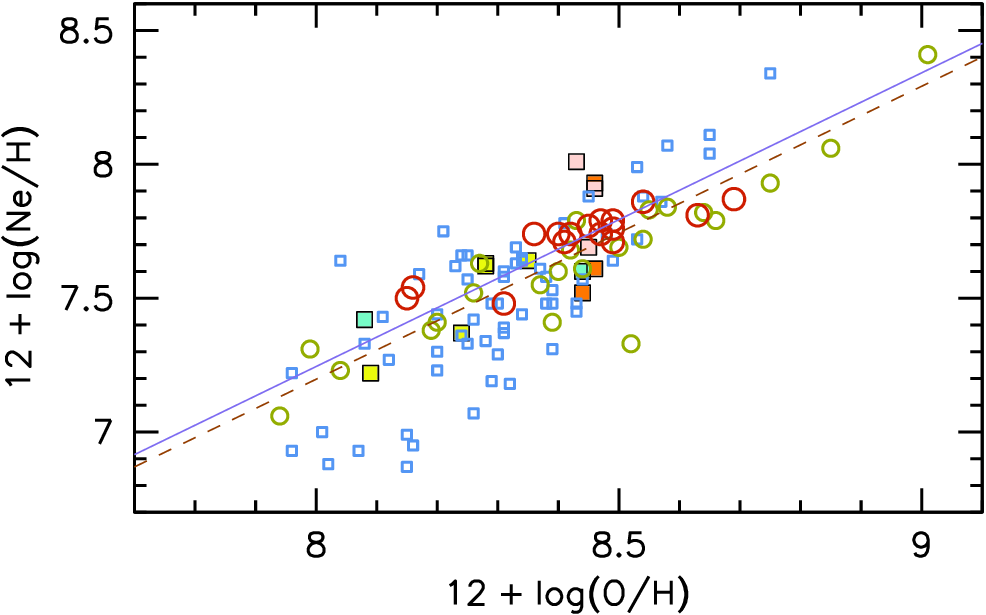}
 \caption{Ne/H  as a function of O/H.  Symbols as in Fig.~\ref{excitation}. The regression to the PNe (our sample + M09) is shown by the dashed line, and the trend found by \citet{Izotov:2006} for more than 400 emission-line  galaxies is shown by the continuous line.
  \label{neo}
}
\end{figure}

Fig.~\ref{neo} shows the values of Ne/H as a function of O/H for PNe and \hii\ regions, with the same symbols as in Fig.~\ref{excitation}.
The regression to the PN data (considering \oiii\lin4363-based abundances only), shown by the dashed line, has a slope of $1.09\pm0.07$, with a correlation coefficient of 0.89. The Ne-O relationship observed by \citet{Izotov:2006} in a sample of star-forming galaxies and blue compact galaxies is shown for reference (continuous line, slope $1.097\pm0.015$).
The increasing deviation  of the RS08 \hii\ region data (blue squares) from the Izotov et al.~line with decreasing O/H has the same nature as the excitation trend we detected in the Ne/O ratio (Fig.~\ref{ne_excitation}), and can be related to an inaccurate ICF(Ne) at low nebular excitation.


\citet{Wang:2008} have pointed out that the Ne/O ratio of the ionized ISM   follows an increasing trend with O/H,
reaching a mean value of $\sim$0.25 in the Milky Way and other metal-rich galaxies, suggesting 
that the Ne enrichment of the ISM is delayed relative to O. The average Ne/O ratio for PNe with \oiii\lin4363 detections in M33 (our sample + M09) is 0.18\,$\pm$\,0.04, comparable with the value observed in the LMC (\citealt{Leisy:2006}), which has a present-day metallicity similar to that of the ISM in M33. 

\begin{figure}
\includegraphics[width=0.47\textwidth]{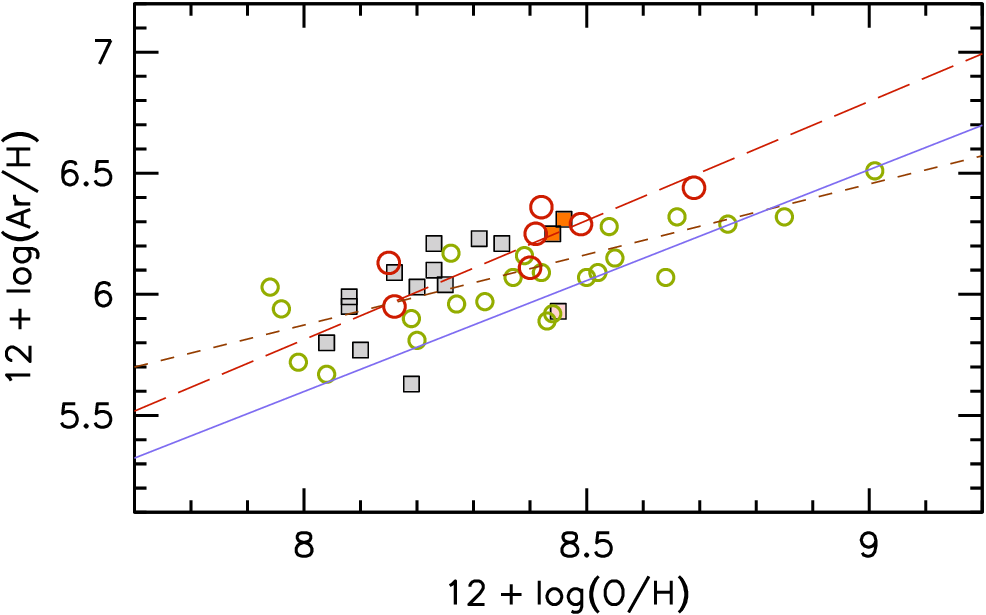}
\caption{The Ar/H abundance ratio as a function of O/H. Symbols as in Fig.~\ref{excitation}. The regression to the PNe (our sample + M09) is shown by the short-dashed line, and the trend found by \citet{Izotov:2006} for more than 400 emission-line  galaxies is shown by the continuous line. The long-dashed line
represents the regression to our PN sample, combined with the \hii\ regions.
  \label{ar_o}
}
\end{figure}

\subsection{Argon and oxygen}\label{argon}
Argon and sulfur are also products of  primary nucleosynthesis in massive stars, and their abundances can be used in alternative to oxygen to characterize the initial composition of PN progenitors, since  they
should not be affected by evolution during the AGB phase (\citealt{Leisy:2006}), at any chemical composition. The main difficulty in the study of argon and sulfur is that observationally it is more challenging than that of oxygen, due either to the weakness of the lines, or the required wavelength coverage. Moreover, the correction for unseen stages of ionization can be important, especially for sulfur (\citealt{Henry:2004}). 
In this work we only consider the abundance of argon, since for sulfur we only have access to the \sii\ lines.

Observations in the Milky Way have shown that argon, like neon, follows oxygen reasonably well in both PNe and \hii\ regions (\citealt{Henry:2004}), although \citet{Stanghellini:2006} failed to find a correlation between Ar and O abundances for PNe, likely as a result of the uncertain Ar abundances.  Our plot of Ar/H \vs\ O/H is shown in Fig.~\ref{ar_o}, where we can see a good correlation ($R_{xy}=0.76$) between the two quantities.
The slope of the regression that considers the whole PN sample (short-dashed line) is $0.58\pm0.09$, significantly lower than  unity, and smaller than the value found for the
star-forming galaxy sample studied by \citet{Izotov:2006}, $0.916\pm0.021$.
For Milky Way PNe, \citet{Henry:2004} obtained a good correlation between Ar/H and O/H, but with an even higher slope of $1.34\pm0.14$. 
In light of the problems with the PN Ar/H abundances mentioned earlier, 
we carried out a regression to our PNe sample and the \hii\ regions combined, since they appear to follow a common relation in the diagram
(long-dashed line).
We find  a slope of $0.98\pm0.27$, consistent with unity, although the error  is quite large. We point out that 
these data lie systematically above the \citet{Izotov:2006} regression line.
We conclude that Ar is rather well correlated with O in both PNe and \hii\ regions in M33, but further studies will be required to 
reduce the uncertainty in the relation between these two quantities for this galaxy.

\begin{figure}
\includegraphics[width=0.47\textwidth]{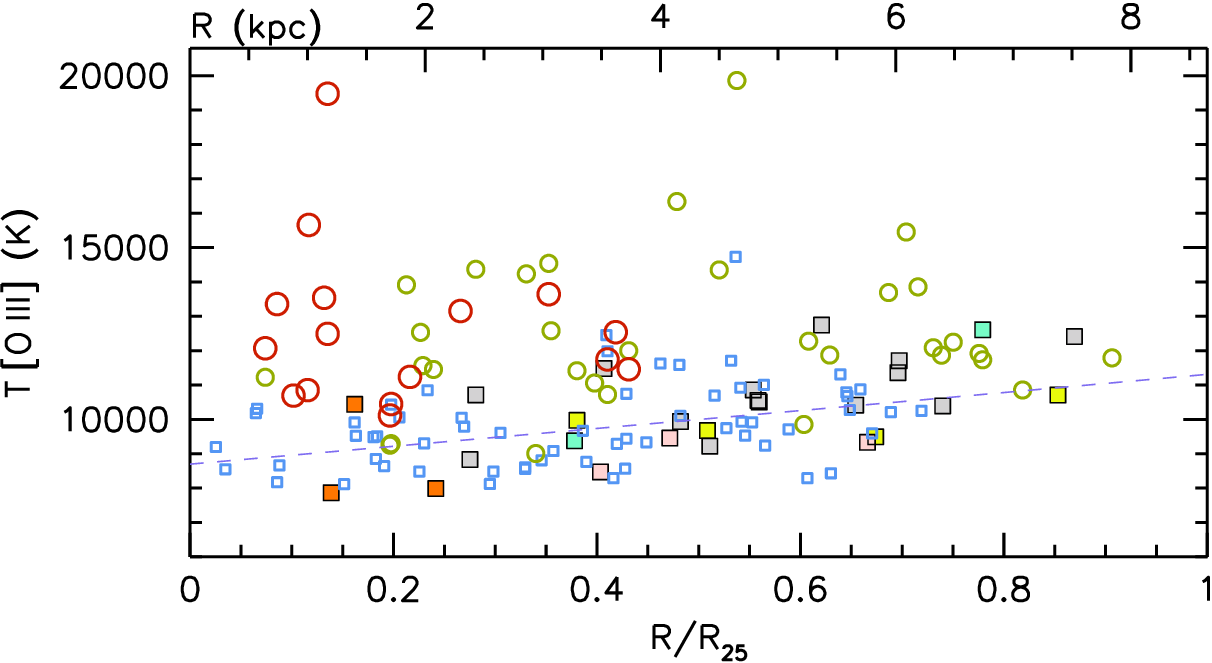}
 \caption{Nebular temperature T\oiii\ as a function of galactocentric distance for PNe and \hii\ regions. Radial distances are given both in terms of the 
 isophotal radius \rtf\ (bottom axis) and in kpc (top axis). Symbols as in Fig.~\ref{excitation}.
The dashed line represents the linear regression to the \hii\ region data.
  \label{radial_te}
}
\end{figure}

\begin{figure*}
\includegraphics[width=0.8\textwidth]{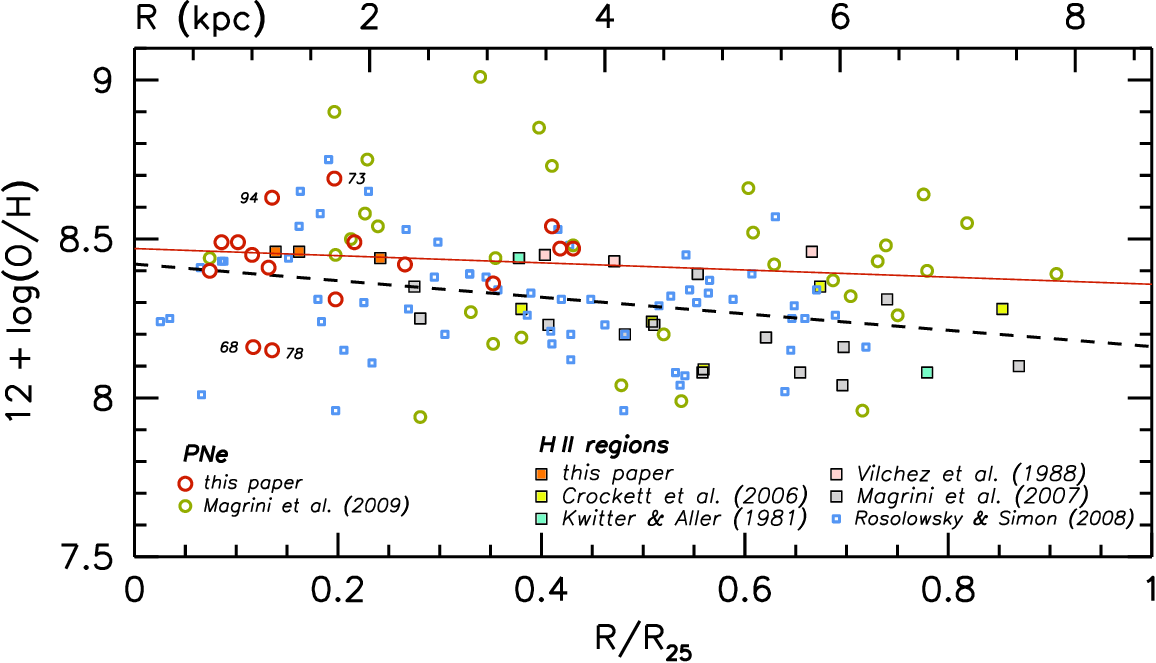}
 \caption{O/H radial gradient for PNe and \hii\ regions. Radial distances are given both in terms of the 
 isophotal radius \rtf\ (bottom axis) and in kpc (top axis). The different symbols used are summarized in the legend.
Linear regressions are shown by the red continuous line (PNe) and the dashed black line (\hii\ regions).
The four planetaries that deviate the most from the PNe regression, PN\,68, 73, 78 and 94, are labeled.
  \label{radial_gas}
}
\end{figure*}

\section{Radial abundance gradients}\label{gradients}

One of our main motivations at the inception of this spectroscopic study of M33 was to test whether 
PNe could provide complementary information on the galactic-scale oxygen abundance gradient in this galaxy and on its possible
temporal evolution.
Despite the fact that several \hii\ region emission line studies have been carried out in M33 throughout the years
(to cite a few: \citealt{Kwitter:1981,Vilchez:1988,Willner:2002,Crockett:2006,Magrini:2007,Rosolowsky:2008,Rubin:2008}), the present day metal content of the central parts of the galaxy remains the most uncertain, due to the difficulties in measuring the required, temperature-sensitive faint auroral lines, as the gas temperature decreases with decreasing galactocentric distance. This is a consequence of the nebular cooling increasing with metallicity.
Fig.~\ref{radial_te} illustrates the systematic change of temperature with distance from the galactic center derived for the \hii\ regions, for which we find a gradient of $300\pm60$~K\,kpc$^{-1}$ (no galactocentric variation in temperature is detected for PNe).

Comparative studies of the abundance gradient in the Milky Way, that include observations of PNe, \hii\ regions and
young stars, provide rather consistent results among the different abundance tracers (\citealt{Maciel:1999,Henry:2004}), but with considerable uncertainties resulting from poorly known distances to individual targets. Chemical elements that are unaffected by
nucleosynthesis in low- and intermediate-mass stars, such as S, Ar and Cl, should preferentially be used, however it appears that O and Ne, which are more easily measured, also trace the abundances of the ISM at the time of formation of the PNe progenitors, since 
their production is very small in the corresponding mass range, except at the lowest metallicities.

\medskip
Our discussion about the Ne/O ratio presented in Section~\ref{neon} confirms
the general finding obtained by other authors in a few nearby galaxies, namely that above the metallicity of the SMC [\oh\,=\,8.1]
the brightest PNe and \hii\ regions have comparable metallicities, as measured by the O/H ratio (\citealt{Richer:1993,Richer:2007}). This allows us to examine
the metallicity gradient in M33 independently of the \hii\ regions, by using PNe. In the comparison, we can also 
include massive supergiant stars, which, owing to their young ages, are also excellent probes of the present day
chemical composition of galaxies. In a recent work on the spiral galaxy NGC~300, \citet{Bresolin:2009a} confirmed that
the analysis of the metal lines of blue supergiants provides an abundance gradient which is in excellent agreement with the one derived from \oiii\lin4363 detections in \hii\ regions, including metallicities near the solar value.
In the case of M33, stellar metallicities for young stars have been published by \citet[B supergiants]{Urbaneja:2005a}
and \citet[A supergiants]{U:2009}. The latter also compared the radial abundance gradient in M33 using different indicators (\hii\ regions, supergiant stars and Cepheids).

Despite the difficulties of obtaining spectra of extragalactic PNe with good signal-to-noise ratios,  the use of PNe presents a couple of advantages over \hii\ regions in abundance studies near the galactic central regions. First, as the metallicity increases at small galactocentric distances, the determination of the decreasing \hii\ region electron temperature becomes very difficult, as a consequence of the more efficient gas cooling that reduces the strengths of the  \oiii\ lines.
In the case of PNe, the excitation provided by the much harder stellar radiation  and the higher electron densities, which reduce the cooling due to collisional de-excitation,  ensure that the detection of \oiii\lin4363 is still feasible. Secondly, PNe are not expected to be affected
by abundance biases at high O/H ratios. \citet{Stasinska:2005} pointed out that in the presence of 
temperature gradients {\em within} \hii\ regions, which are predicted to develop at high metallicity (around and above the solar value), the measured O/H abundances could systematically underestimate the true metal content of the gas. Again, the 
high effective temperature of the ionizing stars of PNe, together with their higher densities, prevents this from occurring.

In the following discussion, we consider optical comparison samples of \hii\ regions taken from various authors: \citet{Kwitter:1981}, \citet{Vilchez:1988}, \citet{Crockett:2006}, \citet{Magrini:2007} and RS08. These works provide \oiii\lin4363 detections 
for the measurement of the electron temperature.
The wavelength coverage differs among the different studies, so that, for example, \nii\ and \ariii\ fluxes are unavailable for the \citet{Crockett:2006} and RS08 samples.
Only nebulae for which the electron temperature could be directly derived from \oiii\lin4363 were retained, leaving a final sample
of 88 nebulae.
To overcome difficulties in the analysis arising from the non-homogeneous nature of the data sets  
and to avoid systematic 
discrepancies that could originate from differences in atomic data, ionization correction factors, etc., we have recalculated the abundances from the published line fluxes using the same procedure employed for our \hii\ region data (see Section~\ref{abundances}).

\subsection{Oxygen}
\begin{figure}
\includegraphics[width=0.47\textwidth]{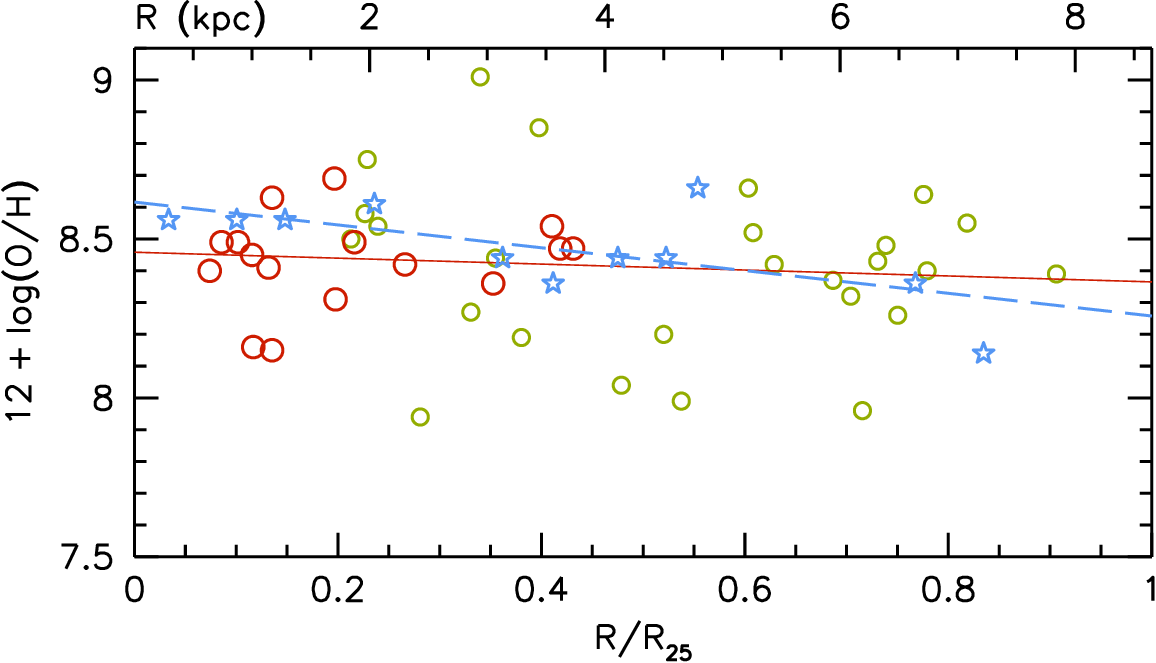}
 \caption{O/H radial gradient for PNe (red and green circles) and B-type supergiants (blue star symbols).
Linear regressions are shown by the red continuous line and the dashed blue line, respectively.
  \label{radial_stars}
}
\end{figure}

The O/H radial behaviour for the PNe having \oiii\lin4363-based abundances is compared with the \hii\ region and the blue supergiant gradients
in Fig.~\ref{radial_gas} and \ref{radial_stars}, respectively. 
We have calculated a linear regression for the combined PNe data (ours and M09), and represent it with the continuous red lines in both figures. 
We have used the weighted least square method described by \citet{Akritas:1996}, to account for the intrinsic scatter in abundance, in addition to uncertainties in the individual abundance values.
The regression yields:\\[-2mm]

\begin{equation}
\rm \oh = 8.47\; (\pm 0.07) - 0.013\; (\pm 0.016)\; R_{\rm kpc} 
\end{equation}

\noindent
The gradient slope that we obtain is compatible within 1-$\sigma$ with both a flat gradient and with the slope 
obtained by M09 ($-0.031\pm0.013$ dex\,kpc$^{-1}$). We note that M09 did not restrict their analysis to objects with auroral line detections, as we do instead. This partly explains the difference in the regression parameters between the two studies. Moreover, our measurements of \oiii\lin4363 in the central 2~kpc of the galaxy provide additional constraints for the gradient slope in the inner disk.

The linear regression to the \hii\ region sample, shown by the dashed line in Fig.~\ref{radial_gas}, is:

\begin{equation}
\rm \oh = 8.42\; (\pm 0.03) - 0.030\; (\pm 0.008)\; R_{\rm kpc} 
\end{equation}

\noindent
Again, this is compatible with the slope of the PNe O/H abundance gradient. Where the PN and \hii\ region samples differ is in the intrinsic scatter, which is 0.20~dex and 0.09~dex, respectively (and easily confirmed qualitatively by inspecting  Fig.~\ref{radial_gas}). However, it is possible that the measurement errors for the M09 data have been under-estimated, because restricting the regression to our new PN sample the scatter is considerably smaller, 0.09~dex (with slope $-0.027\pm0.032$ dex\,kpc$^{-1}$), confirming the suggestion made earlier  that the abundances measured by M09 are less precise than those we measure in our new sample. One can, in principle, expect a larger scatter for PNe than for \hii\ regions, due to the fact that PNe sample a significant age range in the evolution of galaxies. 
We point out that even in our new sample two PNe (PN\,73 and PN\,94)
located in the inner 2 kpc have an approximately solar oxygen abundance (\ohsun8.69, \citealt{Asplund:2009}), deviating 
considerably (+0.2 dex) from the regression line that describes the abundance gradient. At the same time, PN\,68 and PN\,78 lie $\sim$0.3 dex
below the regression.

We conclude that, given the uncertainties, bright PNe and \hii\ regions give a similar picture of the oxygen abundance gradient in M33.
Our PNe sample provides a good number of auroral line-based abundances in the central region of M33, where \hii\ region abundances are more difficult to obtain, due to the nebular cooling. The ionized gas does not suggest deviations from a simple
exponential gradient in O/H \vs\ radius, even in the central region of the galaxy.

The linear regression to the B-type supergiants studied by \citet{Urbaneja:2005a}, shown in Fig.~\ref{radial_stars} by the dashed line, is given by:

\begin{equation}
\rm \oh = 8.61\; (\pm 0.11) - 0.041\; (\pm 0.027)\; R_{\rm kpc} 
\end{equation}

\noindent

The O/H gradient slope from the blue supergiants is slightly steeper than that obtained from the ionized gas, although 
considering the  errors  the stellar gradient is still compatible with the gradient obtained from PNe and \hii\ regions (it is worth noting that  the single star at $R\,=\,7.22$ kpc is largely responsible for the steeper slope).
In conclusion, the oxygen abundance indicators at our disposal (PNe, \hii\ regions and B supergiants) provide a consistent picture of the O/H gradient in M33.

\subsection{Nitrogen}\label{gradient_nitrogen}

\begin{figure}
\includegraphics[width=0.47\textwidth]{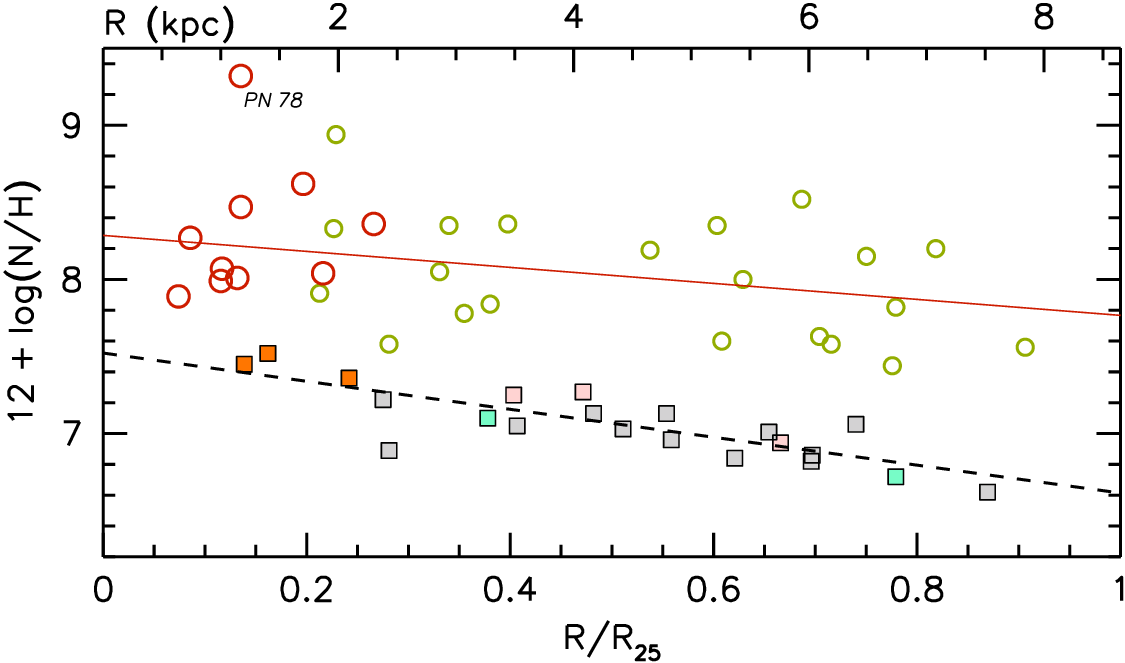}
 \caption{N/H radial gradient for  PNe and \hii\ regions. Symbols and lines as in Fig.~\ref{radial_gas}. PN\,78 was excluded from the
 PN linear regression.
  \label{radial_nh}
}
\end{figure}

The radial N/H abundance gradient for both \hii\ regions and PNe is shown in Fig.~\ref{radial_nh}. The \hii\ regions display 
a well-defined gradient, with a slope $-0.105\pm0.015$ dex\,kpc$^{-1}$, i.e.~significantly steeper than for O/H. 
The slope for the PNe is somewhat shallower ($-0.060\pm0.027$ dex\,kpc$^{-1}$), but with a much larger scatter  (0.31 dex \vs\ 0.09 dex; 
we excluded the high N/H abundance PN\,78 from the fit). 
If we interpret the \hii\ region N/H ratios as the original abundances of the PNe progenitors at birth, the clear offset we see between \hii\ regions and PNe in Fig.~\ref{radial_nh} is a signature of the nitrogen enrichment that took place during the AGB phase of the PN progenitors.

\begin{figure}
\includegraphics[width=0.47\textwidth]{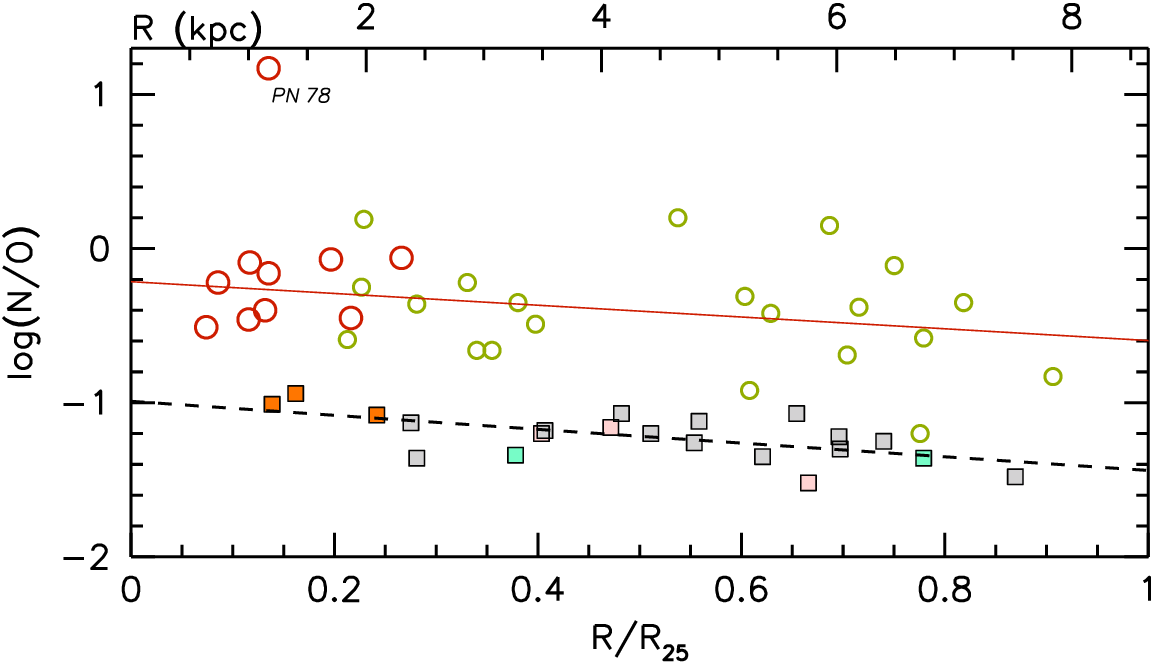}
 \caption{N/O radial gradient for  PNe and \hii\ regions. Symbols and lines as in Fig.~\ref{radial_gas}.  PN\,78 was excluded from the
PN linear regression.
  \label{radial_no}
}
\end{figure}

\begin{figure}
\includegraphics[width=0.47\textwidth]{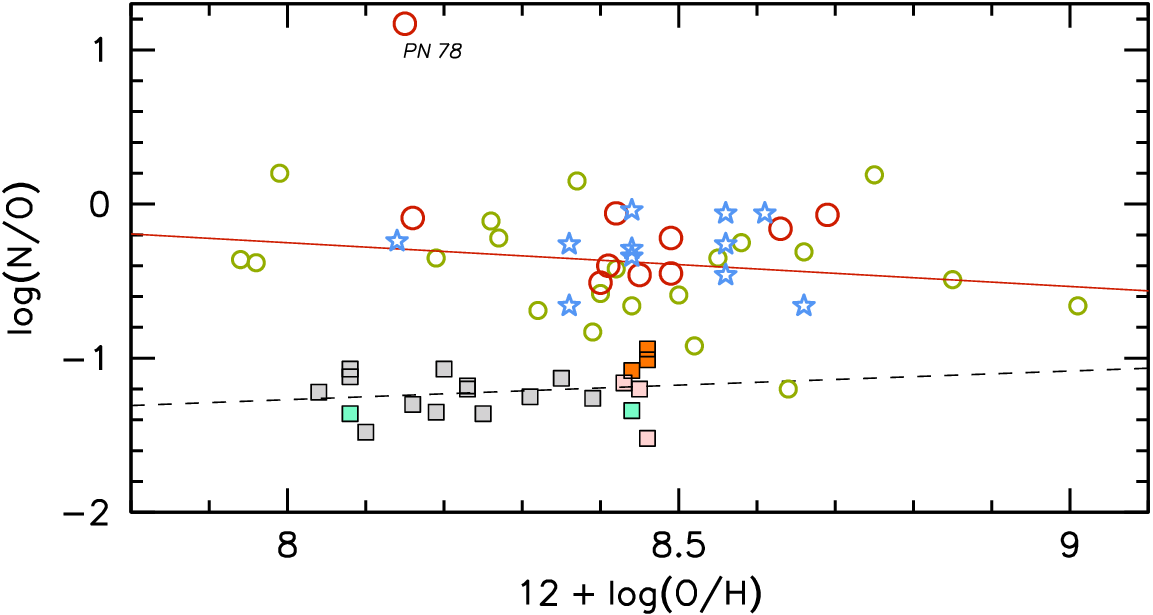}
\caption{N/O  for PNe, \hii\ regions and B-type supergiants as a function of O/H. Symbols and lines as in Fig.~\ref{radial_gas} and \ref{radial_stars}.
PN\,78 was excluded from the PN linear regression.
  \label{oh_no}
}
\end{figure}

The effect of nitrogen enrichment on the N/O radial abundance gradient is displayed in
Fig.~\ref{radial_no}, where we point out  that the two regressions for the \hii\ regions and the PNe are virtually parallel 
(the slopes are $-0.052\pm0.015$ and $-0.044\pm0.025$ dex\,kpc$^{-1}$, respectively). The nitrogen enrichment in the PN progenitors 
is approximately constant, about 0.8 dex, across the metallicity range covered by our samples. 
A similar result was found by \citet{Richer:2007} in the case of the SMC (where the offset between \hii\ regions and PNe was found to be  $\sim$0.7 dex), but in other dwarf irregulars these authors observed that the  N/O ratios span a very wide range ($\sim$ 2 dex) for a given O/H ratio, so it does not seem possible to generalize our result, even though the N/O range in M33 appears relatively smaller. We also note that a few PNe have N/O ratios comparable to those of the \hii\ regions, thus their progenitors did not produce 
large quantities of nitrogen, perhaps as a consequence of their lower masses. Although admittedly the statistics are rather poor, the data in Fig.~\ref{oh_no}, where we plot N/O \vs\ O/H,
 suggest that this is happening preferentially at higher metallicity [\oh\,$>$\,8.5, although this might depend on a poor abundance determination], and that the N/O ratio, and therefore the nitrogen enrichment, is larger at smaller O/H (a similar result was found, among others, by \citealt{Leisy:1996} and \citealt{Stasinska:1998}; but note that, given the errors, the regressions in Fig.~\ref{oh_no} are consistent with flat slopes). 

In Fig.~\ref{oh_no} we also include the N/O ratios calculated for B-type supergiants by \citet[blue star symbols]{Urbaneja:2005a}.
In these post-main sequence stars rotationally-induced mixing of material processed in the stellar core can be brought up to the 
photosphere (\citealt{Maeder:2000}). The N/O abundance ratio, in particular, is heavily enhanced in B supergiants, likely by internal mixing, although additional mechanisms (such as binarity and magnetic fields) can also play an important role (\citealt{Hunter:2008}). 
B supergiants with even larger N/O ratios could be explained by the deep convection occurring during the red supergiant phase, although current theoretical tracks do not extend back to the blue (\citealt{Hunter:2009}).
The stellar data points in Fig.~\ref{oh_no}, compared to the \hii\ regions, show that, interestingly, the N/O ratio
in the B supergiants is approximately the same as for the majority of the PNe, indicating that the nitrogen enrichment 
of present-day evolved massive stars (20-40\,\msun) is comparable to that of the progenitor stars of bright PNe ($\sim$2\,\msun).

\subsection{Neon and argon}

\begin{figure}
\includegraphics[width=0.47\textwidth]{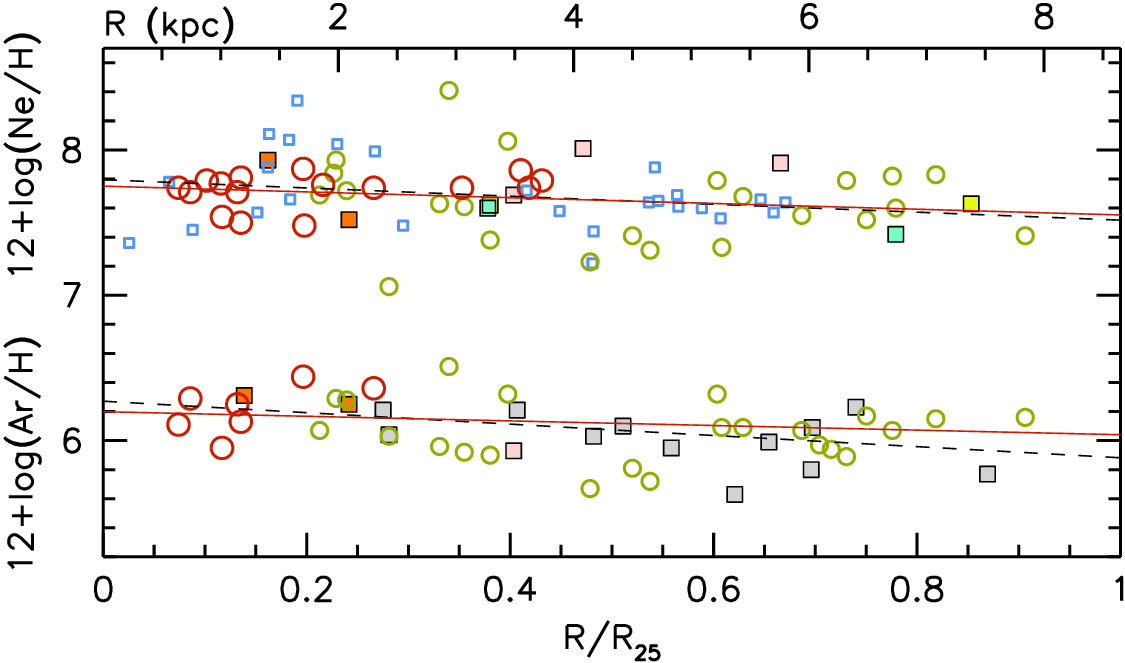}
\caption{Ne/H (top) and Ar/H (bottom) radial gradients for  PNe and \hii\ regions. Symbols and lines as in Fig.~\ref{radial_gas}.
  \label{ne_ar}
}
\end{figure}

The top portion of  Fig.~\ref{ne_ar} displays the radial Ne/H abundance gradient, where we have excluded \hii\ regions 
with \opp/(\opp\,+\,\op)\,$<$\,0.5, since we have evidence that the ICF may be wrong for those. The slopes we obtain for \hii\ regions and PNe, $-0.032\pm0.020$ and $-0.023\pm0.018$ dex\,kpc$^{-1}$, respectively, are 
consistent with each other and with the O/H gradient slope. We must add that for the \hii\ regions the value of the slope is sensitive to the adopted selection in the excitation parameter: if we remove from the fit the 
\hii\ regions for which \opp/(\opp\,+\,\op)\,$<$\,0.6, for example, we obtain a slope of $-0.054\pm0.018$ dex\,kpc$^{-1}$, more in agreement with
the result obtained from Spitzer observations by \citet{Rubin:2008}, $-0.058\pm0.014$ dex\,kpc$^{-1}$.\\

In the bottom plot of Fig.~\ref{ne_ar} we show the radial gradient of the Ar/H abundance ratio for PNe and \hii\ regions. We have omitted the 4 \hii\ regions with 
\opp/(\opp\,+\,\op)\,$>$\,0.6. The gradient slope we show for the PNe is affected by the additional uncertainty represented by the systematic offset we noticed between our sample and the one by M09. This uncertainty is not reflected in the gradient parameter summary of Table~\ref{gradients}.
As the table shows, the slopes we derive for both types of objects agree with the values obtained for the oxygen abundance gradient.

%


\begin{table*}
 \centering
 \begin{minipage}{17.5cm}
  \centering
  \caption{Parameters of linear radial abundance gradients$^1$.}\label{gradients}
  \begin{tabular}{ccccccccc}
  \hline

	&	\multicolumn{4}{c}{PNe}	  & 	\multicolumn{4}{c}{\hii\ regions}	\\[2mm]
			&	Slope & & Correlation & No.\phantom{aa}	& 	Slope & & Correlation & No. \\
Element\phantom{aaaaa}		&	(dex kpc$^{-1}$)	& Intercept	& Coefficient & objects\phantom{aa}	& (dex kpc$^{-1}$)	& Intercept	&	 Coefficient & objects\\

 \hline
 
 O\dotfill	&	$-0.013 \pm 0.016$	&	$8.47 \pm 0.07$	&	$-0.10$	& 48\phantom{aa}  &	$-0.030\pm0.008$	&	$8.42\pm0.03$	&	$-0.34$ & 87\\[1mm]
		
N\dotfill &	$-0.060 \pm 0.027$	&	$8.29 \pm 0.12$	&	$-0.38$	& 36\phantom{aa}  &	$-0.105 \pm 0.015$	&	$7.52 \pm 0.07$ &	$-0.83$ & 21\\[1mm]

 Ar\dotfill	&	$-0.018 \pm 0.014$	&	$6.20 \pm 0.06$	&	$-0.27$	& 36\phantom{aa}  &	$-0.045 \pm 0.016$	&	$6.27 \pm 0.07$ &	$-0.60$ & 15 \\[1mm]

 Ne\dotfill	&	$-0.023 \pm 0.018$	&	$7.75 \pm 0.07$	&	$-0.17$	& 45\phantom{aa}  &	$-0.032\pm0.020$	&	$7.79\pm0.08$	&	$-0.27$ & 35\\


 \hline
\end{tabular}
\end{minipage}
\begin{minipage}{17.5cm}
$^1$Linear relation between 12\,+\,log(X/H) and galactocentric distance in kpc.\\
\end{minipage}

\end{table*}

\subsection{Radial gradient summary}
To summarize our results for this section, we report in Table~\ref{gradients} the parameters of the radial logarithmic abundance gradients for O, N, Ar and Ne. The number of objects used 
for the fits is included.
Given the sources upon which these results are based, it is not surprising that we find a good agreement with the slopes obtained for PNe by M09 (O/H:  $-0.031\pm0.013$ dex\,kpc$^{-1}$; Ne/H: $-0.037\pm0.018$ dex\,kpc$^{-1}$).

It can be seen from Table~\ref{gradients} that, considering PNe and \hii\ regions separately, the slopes for the O, Ar and Ne abundance gradients are consistent with each other. 
In an attempt to reduce the errors for a representative $\alpha$-element gradient slope, we have  combined the abundance data for O, Ar and Ne, separately for PNe and \hii\ regions, adjusting the zero-points by the intercept amounts reported in Table~\ref{gradients}.
A regression to the combined data yields\\[-0.5cm]

$$\rm \frac{d \log(\alpha/H)}{dR} = -0.033\pm0.008\; dex\,kpc^{-1}$$\\[-0.5cm]

\noindent
for \hii\ regions, and

$$\rm \frac{d \log(\alpha/H)}{dR} = -0.018\pm0.010\; dex\,kpc^{-1}$$

\noindent
for PNe. The two slopes agree within the 1-$\sigma$ errors. If there has been an evolution of the gradient with time, it remains undetected with our current uncertainties. Combining the results, we obtain a representative slope of the radial gradient of the $\alpha$-elements in the ISM of M33:

$$\rm ISM: \frac{d \log(\alpha/H)}{dR} = -0.025\pm0.006\; dex\,kpc^{-1}$$

\noindent
We also point out that the M33 central abundances of these $\alpha$-elements are in good agreement between PNe and \hii\ regions. For
oxygen we find a weighted central abundance \oh$_c$\,=\,$8.43\pm0.02$. With the adopted solar ratio [\ohsun8.69, \citealt{Asplund:2009}], this corresponds to 0.55$\pm0.03$ times the solar value. Argon yields a similar result, 0.50$\pm0.08$ solar, while the neon central abundance equals 0.69$\pm0.04$ times the solar value
(note that we prefer to adopt the \citealt{Lodders:2003} solar value for Ar/H, which is 0.15 dex larger than the \citealt{Asplund:2009} value). The slightly discrepant result for Ne could be brought to a better agreement with those for O and Ar
if we adopted a higher solar value for Ne/O, as advocated in a few recent studies (e.g.~\citealt{Wang:2008}).
A solar value (Ne/O)$_\odot$\,=\,$0.22$ would bring agreement between the central O and Ne abundances in M33, relative to solar (the most recent \citealt{Asplund:2009} value is (Ne/O)$_\odot$\,=\,$0.17$).

\section{Conclusion}

Our work highlights the importance of deriving auroral line-based chemical abundances for large samples of planetary nebulae and \hii\ regions in nearby galaxies. These observations are essential to learn more about systematic errors in nebular diagnostics, the time evolution of the chemical composition of star-forming galaxies, and the nucleosynthesis of intermediate-mass stars. Key targets for these studies are nearby irregular and spiral galaxies, where the spectroscopy of faint diagnostic lines can be carried out with the present generation of telescopes and instruments. 

We have obtained new deep optical spectrophotometry of 16 planetary nebulae in M33, mostly located in the central two kpc of the galaxy. For this sample we have derived electron temperatures and chemical abundances from the detection of the \oiii\lin4363 line. After combining our abundances with those we obtained with the same direct method from the emission line observations of \citet[32 PNe]{Magrini:2009}, we have examined the behavior of nitrogen, neon, oxygen and argon in relation to each other in PNe and \hii\ regions, and as a function of galactocentric distances. 

The N/O and He/H abundance ratios of most of the PNe of our new sample, which comprises objects in the brightest 1.5 magnitudes of the \oiii\lin5007 luminosity distribution, lead to a Type~I classification, which, at least in the Milky Way, corresponds to young objects with relatively massive progenitors. We have found one object, PN\,78, with an extreme nitrogen abundance, accompanied
by a large helium content. 

We confirm the good correlation between Ne/H and O/H abundances for PNe in M33, as expected from the primary nature of these two elements. We also find a linear correlation between argon and oxygen, but we point out that this is somewhat uncertain, due possibly to observational errors. 
The Ar/O ratio is consistent with the value found in \hii\ regions and in the Sun. For Ne/O, we find a mean value for PNe of 0.20, which is 18\% larger
than the \citet{Asplund:2009} value, but in agreement with the Ne/O \vs~O/H trend found by \citet{Wang:2008}.
These result lend support to the idea that at the metallicities of the PNe analyzed in M33 these three
$\alpha$-elements (O, Ar and Ne) trace the abundance of the progenitor stars at the moment of their birth, and in particular that neon and oxygen have not been modified during the dredge-up process taking place during the AGB phase.

We find no significant abundance offset between PNe and \hii\ regions at any  galactocentric radius, despite the fact that they represent different age groups in the evolution of the galaxy. The combination of the PN and \hii\ region auroral-line abundances provide information on the shape of the oxygen abundance gradient in the central few kpc of M33.  
We obtain an $\alpha$-element abundance gradient of the ISM in M33 with a slope of $-0.025 \pm 0.006$ dex\,kpc$^{-1}$, and the slopes obtained separately for \hii\ regions and PNe agree, within the current uncertainties.
This is consistent with the conclusion by M09 that the disk of M33 has not experienced a significant chemical enrichment of its interstellar medium in the past few Gyr. Moreover, despite the large dispersion in O/H abundances for both PNe and \hii\ regions, we find no indication that 
towards the center of the galaxy PN values are systematically larger than \hii\ region ones, as would be the case if \hii\ region abundances were biased towards lower values, and as could be the case if O/H in the central parts of M33 were really high (super-solar). This confirms what B supergiants already indicated for this galaxy.

The growing amount of data on abundance tracers in M33 in the past few years has raised new questions regarding the present-day abundance gradient in this galaxy. For example, what is the origin of the large intrinsic dispersion of \hii\ region O/H abundances (also displayed by the PNe data shown here)
at a given radius discovered by RS08? We found no correlation of the abundance residuals, as measured with respect to the mean gradient, with a number of parameters, including nebular luminosity, equivalent width of the H$\beta$ line, line strengths, gas density, or position in the galaxy disk.
An analysis of additional data acquired as part of the M33 Metallicity Project (RS08) will hopefully  shed some light on this issue.
Moreover, the study of A supergiants by \citet{U:2009} provided metallicities near the galaxy center that lie  above the solar value, and their
derived slope, corrected to account for our adopted distance, is $-0.084\pm0.010$ dex\,kpc$^{-1}$. While it is true that the comparison between 
stars and ionized gas is complicated by the fact that A supergiant metallicities
refer mostly to the iron content (rather than that of the $\alpha$-elements), it is striking, as also pointed out by \citet{U:2009}, that the stellar metallicities are consistently near or above the solar value close to the center of M33, while the \hii\ regions samples includes objects with O/H ratios that extend down to 20\% of the solar value. Future studies of the oxygen content of A and B supergiants in the central few kpc of M33 will help to 
directly compare stellar and nebular abundances, and perhaps better understand the nature of the abundance dispersion.

\bigskip
\bigskip
\noindent
FB would like to thank the Observatoire de Paris (Meudon) and the
Instituto de Astrof\'{\i}sica de Andaluc\'{\i}a (Granada) for hospitality during visits in which part of this work was carried out.
FB gratefully acknowledges the support from the National Science Foundation grant AST-0707911. ER's work is supported by a Discovery Grant from NSERC of Canada.
We thank L. Magrini, for providing us with data in advance of publication, and the staff of the Subaru telescope, in particular T. Hattori, for the support provided for this project.
We are grateful to the referee, Jeremy Walsh, for providing comments that helped us to improve the manuscript.

\bibliography{m33}


\end{document}